\documentclass[aps,pre,twocolumn,longbibliography]{revtex4-1}
\usepackage{amssymb}
\usepackage{bm}
\usepackage{graphicx}

\begin{document}

\title{Rotating Rayleigh-Taylor turbulence}

\author{G. Boffetta}
\affiliation{Department of Physics and INFN, University of Torino, 
via P. Giuria 1, 10125 Torino, Italy}
\author{A. Mazzino}
\affiliation{Department of Civil, Chemical and Environmental Engineering,
INFN and CINFAI,
University of Genova, Via Montallegro 1, 16145 Genova, Italy}
\author{S. Musacchio}
\affiliation{Universit\'e de Nice Sophia Antipolis, CNRS, 
Laboratoire J.A. Dieudonn\'e, UMR 7351, 06100 Nice, France} 
\date{\today}

\begin{abstract}
The turbulent Rayleigh--Taylor system in a rotating reference frame is 
investigated by direct numerical simulations within the Oberbeck-Boussinesq
approximation.
On the basis of theoretical arguments, supported by our simulations,
we show that the Rossby number decreases in time, and therefore
the Coriolis force becomes more important as the system evolves
and produces many effects on Rayleigh--Taylor turbulence.  
We find that rotation reduces the intensity of turbulent velocity fluctuations 
and therefore the growth rate of the temperature mixing layer.
Moreover, in presence of rotation the conversion of potential energy 
into turbulent kinetic energy is found to be less effective 
and the efficiency of the heat transfer is reduced. 
Finally, during the evolution of the mixing layer 
we observe the development of a cyclone-anticyclone 
asymmetry.
\end{abstract}
\maketitle 

\section{Introduction}
\label{sec:1} 

In natural fluid systems, 
the interface between regions of different density 
is subject to the well-known Rayleigh-Taylor (RT) instability
when the heavier fluid is pushed against the lighter one 
(see, e.g., 
\cite{chandrasekhar1961hydrodynamics,sharp1984overview,abarzhi2010review} 
for reviews on the subject). 
In the late stage, this instability evolves in a fully-developed
turbulent flow  \cite{sharp1984overview,abarzhi2010review,boffetta2017}.
Because of the fact that unstably stratified interfaces are
ubiquitous in nature, fluid mixing induced by RT instability and turbulence
characterizes many systems, over a huge interval of spatial and
temporal scales.
Among the possible examples, we mention astrophysical systems,
in relation to flame acceleration in type Ia supernova
\cite{hillebrandt2000type,bell2004direct};
geological systems, where RT instability has been invoked to explain
the initiation and evolution of polydiapirs (domes-in-domes)
\cite{weinberg1992polydiapirs};  atmospheric physics,
where RT instability has been invoked as a possible explanation for 
the formation of the mammatus clouds \cite{agee1975some}.

Several experimental 
\cite{read1984experimental,dalziel1993rayleigh,dimonte1996turbulent,banerjee2006statistically,andrews2010small,banerjee2010detailed} 
and numerical \cite{young2001on,dimonte2004comparative,cabot2006reynolds,vladimirova2009self,matsumoto2009anomalous,biferale2010high} works have investigated 
the evolution of RT instability and turbulence in different physical conditions.
In the limit of small density variations, the simplest mathematical 
description is provided by the Oberbeck-Boussinesq approximation
of two incompressible, miscible fluids.
Under these conditions, the seminal paper
\cite{chertkov2003phenomenology} 
has shown that three-dimensional RT turbulence can be described
as a time-dependent hydrodynamical turbulent flow 
with a gravity-induced forcing mechanism. 

The RT instability and mixing can be significantly affected by the 
presence of other physical mechanisms, beside gravity, acting on the flow, 
including compressibility \cite{mitchner1964rayleigh,
scagliarini2010lattice}, viscoelasticity \cite{boffetta2010polymer},
surface tension \cite{chertkov2005effects} and
physical confinement \cite{lawrie2011turbulent,boffetta2012bolgiano}.
An important instance is the case in which the flow is embedded in a 
rotating reference frame, which causes the emergence of the Coriolis 
force in the ruling equations.
Previous works have shown that the Coriolis force has a stabilizing
effect on the RT instability, both in the linear stage
\cite{chandrasekhar1961hydrodynamics}
and in the weakly nonlinear stage of the perturbation evolution
\cite{tao2013nonlinear,carnevale2002rotational,baldwin2015inhibition}. 

In this study we focus on the effects of rotation 
on the fully developed RT turbulence. 
We discuss theoretical arguments based on dimensional scaling analysis 
which predict that the Rossby number decreases in time 
during the growth of the mixing layer. 
The Coriolis force is therefore expected 
to become dominant as the system evolves. 
This prediction is confirmed by our numerical simulations, 
which reveals that the turbulent RT system is strongly affected by rotation.
In particular, we find that rotation 
reduces the efficiency of the heat transfer 
as well as the conversion of potential energy 
into turbulent kinetic energy. 
This causes a reduction of the intensity of turbulent 
velocity fluctuations and slows down the growth of the mixing layer. 
Interestingly, rotation allows the velocity fluctuations to 
extend outside the temperature mixing layer. 
We also observe the development of a cyclone-anticyclone 
asymmetry which is weakly dependent on the evolution of the mixing layer. 

The paper is organized as follow. In Sec.~\ref{sec:2} we introduce the rotating
RT system under the Oberbeck-Boussinesq framework and we present mean field
(scaling) arguments through which the role of rotation can be inferred.
Numerical results from high-resolution DNS of the rotating RT system are
presented and discussed in Sec.~\ref{sec:3} 
concerning the effects of rotation on 
i) the mixing-layer growth,  
ii) the heat transfer, 
iii) the energy spectra, 
and, finally, 
iv) the symmetry breaking between cyclones and anticyclones. 
Final discussions and perspectives are drawn in Sec.~\ref{sec:4}

\section{Equation of motion and theoretical considerations}
\label{sec:2}
We consider miscible Rayleigh-Taylor turbulence generated at the 
interface of two layers of fluid with small temperature (density) 
difference. The initial (at $t=0$) interface is the plane $z=0$, 
perpendicular to gravity ${\bm g}=(0,0,-g)$ and to the uniform rotation
${\bm \Omega}=(0,0,\Omega)$.
The dynamics is ruled by the Boussinesq equations for an 
incompressible (${\bm u}({\bm x},0)=0$) velocity field 
${\bm u}({\bm x},t)=(u,v,w)$
coupled with a temperature field $T({\bm x},t)$
\begin{equation}
{\partial {\bm u} \over \partial t}+ {\bm u} \cdot {\bm \nabla} {\bm u} + 
2 {\bm \Omega} \times {\bm u} = - {\bm \nabla} p + \nu \nabla^2 {\bm u} -
\beta {\bm g} T
\label{eq:2.1}
\end{equation}
\begin{equation}
{\partial T \over \partial t}+ {\bm u} \cdot {\bm \nabla} T =
\kappa \nabla^2 T
\label{eq:2.2}
\end{equation}
The reference temperature is set to $T=0$,
$\nu$ and $\kappa$ are the kinematic viscosity and diffusivity 
respectively and $\beta$ is the thermal expansion coefficient.
The initial conditions are ${\bm u}({\bm x},0)=0$, 
$T({\bm x},0)=-(\theta_0/2)sgn(z)$ where $\theta_0$ 
is the temperature jump 
at the interface at $z=0$,
which fixes the Atwood number $A=\beta \theta_0/2$.

As turbulence develops from the instability at the interface, turbulent
kinetic energy in the mixing layer $E=(1/2) \langle | {\bm u}|^2 \rangle$ is 
produced at the expense of potential energy 
$P=-\beta g \langle z T \rangle$ (brackets represent integral over the
physical domain of volume $L_x \times L_y \times L_z$). 
The energy balance is written from 
(\ref{eq:2.1}-\ref{eq:2.2}) as
\begin{equation}
-{dP \over dt} = \beta g \langle w T \rangle = {dE \over dt} +
\varepsilon_{\nu}+\beta g {\kappa \theta_0 \over 2 L_z}
\label{eq:2.3}
\end{equation}
where $\varepsilon_{\nu}=\nu \langle ({\bm \nabla} {\bm u})^2 \rangle$
is the viscous energy dissipation and the last term is negligible in the 
limit of small diffusivity. We define the {\it efficiency} $\Sigma$
of turbulent generation as the ratio of the kinetic energy production
over potential energy consumption,
$\Sigma=-{dE/dt \over dP/dt}$.
We anticipate that, although the Coriolis force 
$2 {\bm \Omega} \times {\bm u}$ does not enter
directly in the energy balance (\ref{eq:2.3}) it does affect the
efficiency of the process which, for $\Omega>0$ is reduced with 
respect to the value $\Sigma=0.5$ observed for $\Omega=0$
\cite{boffetta2010statistics}.

The development of the RT instability produces a mixing zone of 
width $h(t)$ which grows in time. 
Dimensional arguments derived in absence of
rotation~\cite{boffetta2017} 
predicts that in the turbulent regime the mixing layer grows as
$h(t) \simeq A g t^2$ and the typical velocity fluctuations 
inside the mixing layer grow as $u \simeq A g t$ 
(here $u$ indicates an arbitrary component of the velocity). 

Assuming that the dimensional scalings hold also in the rotating case, 
one obtains the prediction that the dimensionless Rossby number 
$Ro=w_{rms}/(2 \Omega h)$, which measures the relative strength of 
the inertial forces to the Coriolis forces,
should decrease in time as $Ro \simeq 1/(\Omega t)$. 
In particular, in the case of a weak rotation rate $\Omega$ 
even if the effect of rotation is negligible at the initial times 
(ensuring the validity of the dimensional scalings), 
it becomes more important and eventually competes with
the inertial and buoyancy forces for $t \gtrsim 1/\Omega$.
For large $\Omega$, this time can be sufficiently small for the rotation 
to affect already the linear (or quasi-linear) phase of the instability
\cite{carnevale2002rotational,baldwin2015inhibition}.

\begin{figure}[h!]
\includegraphics[width=0.45\columnwidth]{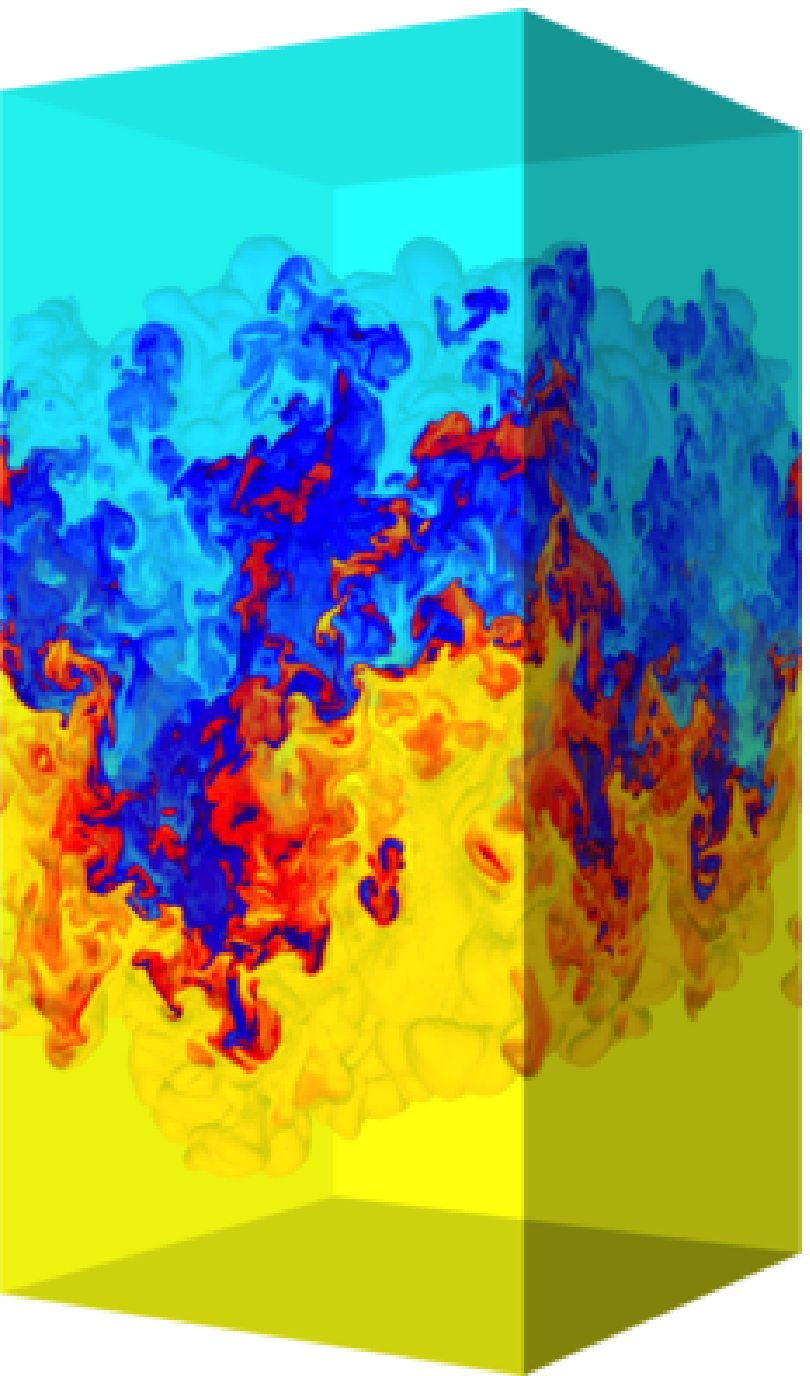}
\includegraphics[width=0.45\columnwidth]{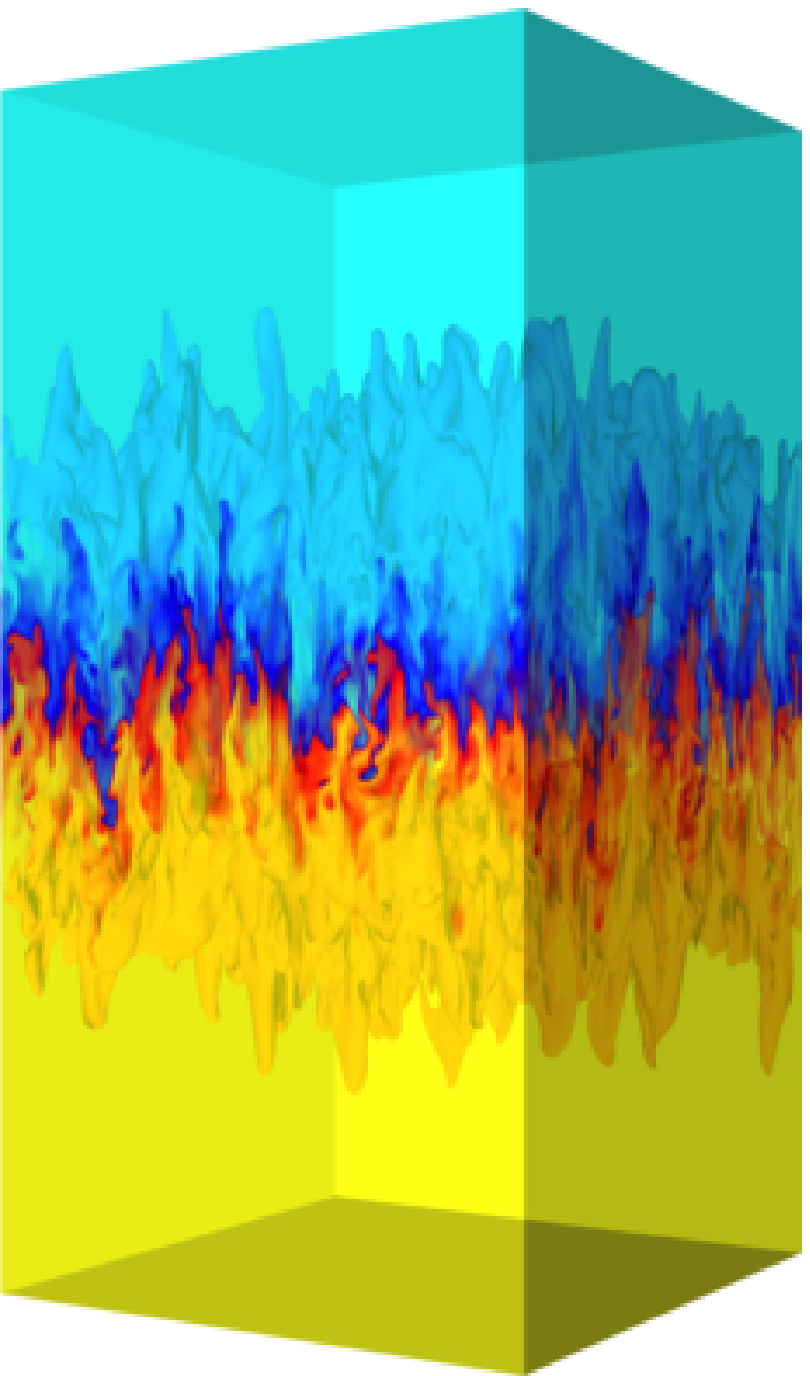}
\caption{Snapshots of the temperature field at time $t=5.2 \tau$
for two simulations with
$\Omega \tau=0$ (left) and $\Omega \tau=10$ (right) starting from the
same initial conditions. Yellow (blue) represents hot (cold), light (heavy)
fluid.}
\label{fig2.1}
\end{figure}

We performed direct numerical simulations (DNS) of the system of equations
(\ref{eq:2.1}-\ref{eq:2.2}) in a periodic domain of size
$L_x \times L_y \times L_z$ with $L_y=L_x$ and $L_z=4 L_x$ by means of a fully 
parallel pseudo-spectral code at resolution $512 \times 512 \times 2048$ 
at different values of rotation $\Omega$. Time evolution is ruled by a 
second-order Runge-Kutta scheme with explicit integration of the linear
terms. For all runs we have $Ag=0.25$, $\theta_0=1$ and $Pr=\nu/\kappa=1$.
Viscosity is sufficiently large to resolve small scales 
($k_{max} \eta \ge 1.3$ at the end of the simulations) 
and the results are made dimensionless 
with the box size $L_x$, the characteristic time $\tau=(L_x/A g)^{1/2}$
and the characteristic velocity $U=L_x/\tau$.

Rayleigh-Taylor instability is seeded by adding to the initial density 
field a $10 \%$ of white noise in a thin layer around the unstable interface
at $z=0$.  The same realization of the random perturbation is
used for all the simulations at different rotation.
The simulation is stopped before the mixing layer reaches the dimension
of the vertical scale $L_z$. 
We remark that the average quantities in (\ref{eq:2.3}) are defined by the
integral over the space and therefore their time dependence has an additional
$t^2$ factor due to the growth of the support of the integral given
by the mixing layer width $h(t)$.

Figure~\ref{fig2.1} shows two snapshots of the temperature field for two
simulations at $\Omega \tau=0$ and $\Omega \tau=10$. Rotation produces 
clear qualitative effects on the temperature field: Thermal plumes are 
more coherent and elongated in the vertical direction while their vertical
velocity is reduced.

\section{Numerical results}
\label{sec:3}
\begin{figure}[h!]
\includegraphics[width=\columnwidth]{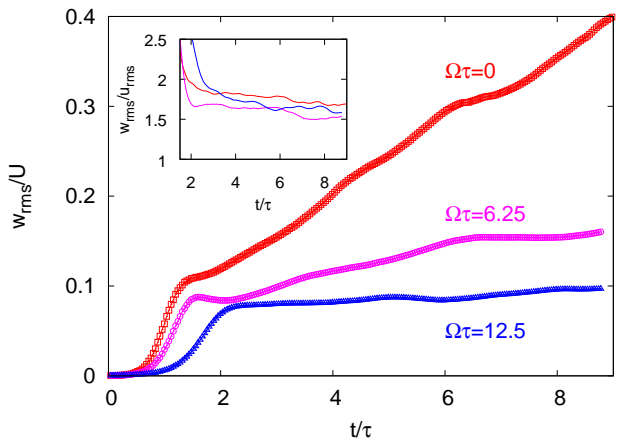}
\caption{
Vertical velocity fluctuations $w_{rms}$ in the mixing layer 
as a as a function
of time for three different rotations, $\Omega \tau=0$ (red squares),
$\Omega \tau=6.25$ (pink circles) and $\Omega \tau=12.5$ (blue triangles).
Inset: ratio of vertical to horizontal velocity fluctuations as a function
of time, for the same values of rotation.}
\label{fig3.1}
\end{figure}

One effect of rotation is to reduce the growth of
the mixing layer, a feature which is evident at a qualitative 
level from Fig.~\ref{fig2.1}.
More quantitatively, this reduction is due to a suppression of the vertical 
velocity fluctuations, measured by $w_{rms}$ and shown in Fig.~\ref{fig3.1}. 
After the instability, for $t> \tau$, vertical velocity fluctuations 
in the non-rotating case $\Omega=0$
are observed to grow linearly in time, in agreement with the dimensional
prediction $w_{rms} \simeq A g t$. 
By increasing the rotation velocity, 
the growth of $w_{rms}$ is strongly depleted and for $\Omega \tau=12.5$ 
vertical velocities almost saturates at a constant value.
The inset of Fig.~\ref{fig3.1} shows the ratio $w_{rms}/u_{rms}$
of the vertical to the horizontal velocity fluctuations
which is found to be weakly dependent on $\Omega$. Therefore also horizontal
velocities are reduced and the degree of large-scale anisotropy
in the velocity field is not strongly affected by rotation. 

\begin{figure}[h!]
\includegraphics[width=\columnwidth]{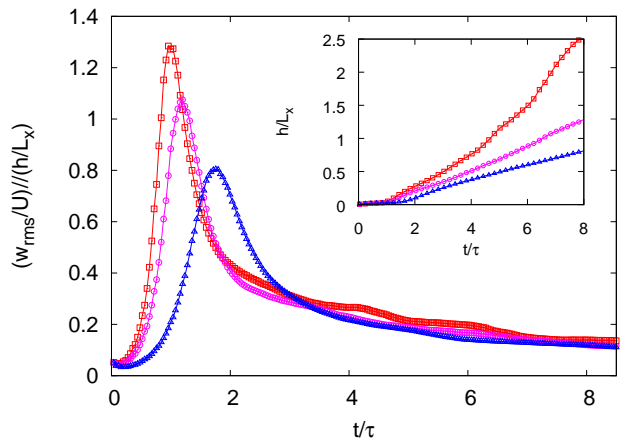}
\caption{
Ratio of the vertical velocity fluctuations $w_{rms}$ to the width 
of the mixing layer $h$ as a as a function
of time for three different rotations, $\Omega \tau=0$ (red squares),
$\Omega \tau=6.25$ (pink circles) and $\Omega \tau=12.5$ (blue triangles).
Inset: time evolution of the mixing layer width $h$ defined in terms of 
a threshold $r \theta_0$ in the temperature profile, with $r=0.9$.}
\label{fig3.1b}
\end{figure}

One consequence of the reduction of velocity fluctuations is the slowing
down of the growth of the mixing layer in presence of rotation.
Different definitions of the mixing layer width $h$ have been proposed,
based on either threshold values or integral quantities \cite{dalziel1999self}.
In the inset of Fig.~\ref{fig3.1b} we plot the time evolution of $h$ defined 
in terms of the threshold value $z=h/2$ at which the temperature 
profile $\overline{T}(z)$ reaches a fraction $r$ of the maximum value,
i.e. $\overline{T}(\pm h/2)=\mp r \theta_0$ 
\cite{dalziel1999self,boffetta2010nonlinear}
(the overline indicates average over the horizontal planes $L_x \times L_y$).
We observe that rotation indeed induces a strong suppression of the growth
of $h(t)$.

A remarkable result shown in Fig.~\ref{fig3.1b} is that
the ratio $w_{rms}/h$ is almost independent on $\Omega$ in the
late turbulent phase. The natural interpretation is that, since
the mixing layer is produced by the velocity fluctuations, $h$ is 
essentially given by the integral in time of $w_{rms}$ and their ratio 
becomes $\Omega$-independent. 

\begin{figure}[h!]
\includegraphics[width=\columnwidth]{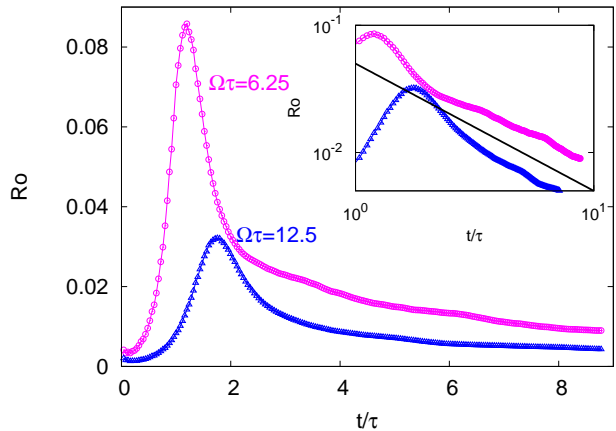}
\caption{
Rossby number $Ro=w_{rms}/(2 \Omega h)$ as a function of time for 
the two rotations 
$\Omega \tau=6.25$ (pink circles) and $\Omega \tau=12.5$ (blue triangles).
Inset: the same quantities in log-log plot. The black line represents 
$t^{-1}$ scaling.}
\label{fig3.1c}
\end{figure}

In Figure ~\ref{fig3.1c} we plot the temporal evolution of the Rossby
number defined as $Ro = w_{rms}/(2\Omega h)$.
Even though in the strongly rotating case the $w_{rms}$ and $h$
might deviate from the scaling expected for $\Omega=0$, 
we find that the Rossby number, 
which is proportional to their ratio, 
decreases with time 
close to the dimensional prediction $Ro \simeq 1/(\Omega t)$.

\begin{figure}[h!]
\includegraphics[width=\columnwidth]{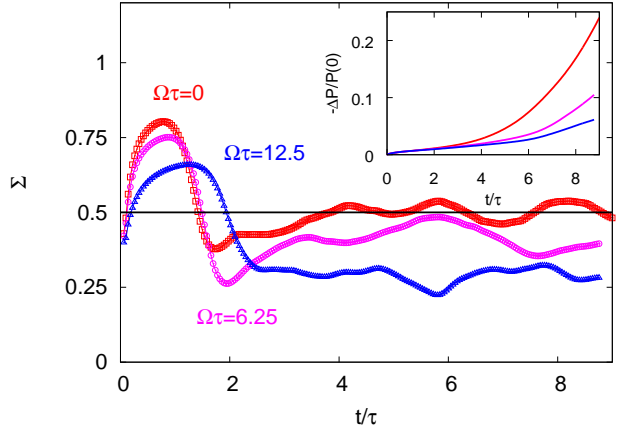}
\caption{
Efficiency of kinetic energy production 
$\Sigma=-(dE/dt)/(dP/dt)$ as a function
of time for three different rotations, $\Omega \tau=0$ (red squares),
$\Omega \tau=6.25$ (pink circles) and $\Omega \tau=12.5$ (blue triangles).
Inset: potential energy loss $-\Delta P(t)=P(0)-P(t)$
for the same three values of rotations (from top to bottom).}
\label{fig3.2}
\end{figure}

Since vertical velocity fluctuations are responsible for the
conversion of potential energy into kinetic energy, 
the potential energy loss $\Delta P(t)=P(0)-P(t)$ 
is reduced for $\Omega>0$, as shown in Fig.~\ref{fig3.2}.
We find that also the efficiency of the process is reduced 
by rotation: For $\Omega=0$ we have $\Sigma \simeq 0.5$, 
i.e. about one half of potential energy is converted into 
large-scale kinetic energy while the other half feeds the turbulent cascade 
and is eventually dissipated by viscosity \cite{boffetta2010statistics}.
For $\Omega>0$ the efficiency is reduced and we find $\Sigma \simeq 0.3$ for
the case $\Omega \tau=12.5$, as shown in Fig.~\ref{fig3.2}, i.e. in this
case more potential energy is transferred to small scales and dissipated.

\subsection{Integral lengths}
Rotation produces different effects on the temperature and on
the velocity fields in RT convection.
To investigate this issue, we introduce a set of integral 
lengths based on the vertical profile  
$f^{(u,w,T)}(z)$ of velocity or temperature variance, defined as
$f^{(u)}(z)=\overline{u^2}(z)$, $f^{(w)}(z)=\overline{w^2}(z)$,
$f^{(T)}(z)=\overline{T^2}(z)-\overline{T}(z)^2$:
\begin{equation} 
L^{(u,w,T)} = \left[ \frac{\int f^{(u,w,T)}(z) z^2 dz }
{\int f^{(u,w,T)}(z) dz} \right]^{1/2}
\label{eq:3.1}
\end{equation} 
where the integral are defined over the whole domain $[0,L_z]$.
The advantage of this definition with respect to the usual $h(t)$,
is that it provides information on
the extension of the mixing layer both in terms of (horizontal/vertical)
velocity and temperature fluctuations. 

\begin{figure}[h!]
\includegraphics[width=\columnwidth]{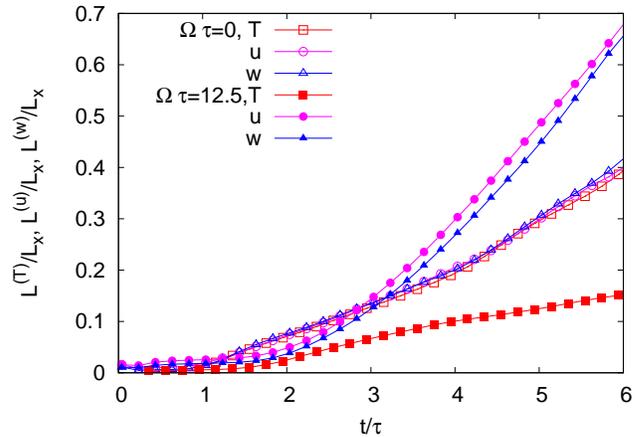}
\caption{
Integral lengths based on the 
standard deviation of the fluctuation profile of
temperature $L^{(T)}$ (red squares), horizontal velocity $L^{(u)}$ 
(pink circles) and vertical velocity $L^{(w)}$ (blue triangles) without 
rotation (empty symbols) and for $\Omega \tau=12.5$ (filled symbols).}
\label{fig3.3}
\end{figure}

The evolution of $L^{(u,w,T)}$ is shown in Fig.~\ref{fig3.3} 
for two values of $\Omega$. In the absence of rotation the three
integral lengths are approximately the same, indicating that
temperature and velocity fluctuations are tightly associated 
within the mixing layer. On the contrary, for $\Omega \tau=12.5$ we
observe that while $L^{(T)}$ is reduced (which is already
clear from Fig.~\ref{fig2.1}),
both $L^{(u)}$ and $L^{(w)}$ increase, approximately by the same amount. 
This is a manifestation of the Taylor-Proudman phenomenon of
bidimensionalization of the velocity field in presence of rotation
\cite{tritton1988physical}
which, becoming almost independent on $z$, extends outside the mixing
layer region defined by temperature fluctuations.

\begin{figure}[h!]
\includegraphics[width=\columnwidth]{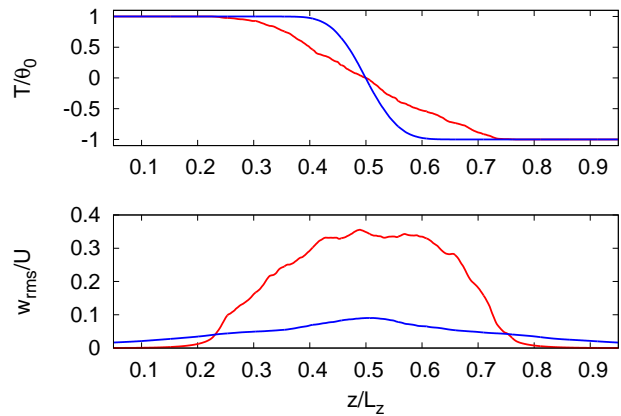}
\caption{
Vertical profiles of mean temperature 
$\overline{T}(z)$ (upper plot) and of vertical velocity
$\overline{w}_{rms}(z)=\overline{w^2}(z)^{1/2}$ for the 
simulation with $\Omega \tau=0$ (red lines) and
$\Omega \tau=12.5$ (blue lines) at time $t=6 \tau$.}
\label{fig3.3b}
\end{figure}

To investigate more in details this issue, in Fig.~\ref{fig3.3b} we
plot the temperature profile $\overline{T}$ and the vertical velocity
profile $\overline{w}_{rms}=\overline{w^2}^{1/2}$ at a late time.
It is evident that while in the case $\Omega=0$ the two profiles 
extend over comparable regions, for the case $\Omega \tau=12.5$
the vertical extension of temperature fluctuations is reduced 
and at the same time the vertical extension of velocity fluctuations
is increased.
Summarizing, our findings show that in presence of rotation there is not 
a unique scale characterizing the mixing layer 
as velocity and temperature fluctuations define different vertical scales. 

\subsection{Heat transfer}
The efficiency of the transfer of heat in turbulent convection is 
quantified by the Nusselt number 
$Nu \equiv \langle w T \rangle h/(\kappa \theta_0)$ 
(ratio of the convective to conductive heat transfer) while turbulence 
intensity is provided by the Reynolds number $Re \equiv |{\bm u}|_{rms} h/\nu$. 
These numbers depends on the dimensionless temperature difference 
quantified by the Rayleigh number 
$Ra \equiv \beta g \theta_0 h^3/(\nu \kappa)$. 
High resolution direct numerical simulations have shown that in 
Rayleigh-Taylor turbulence $Nu$ and $Re$ displays the so-called
ultimate state regime according to which 
$Nu \simeq Ra^{1/2} Pr^{1/2}$ and $Re \simeq Ra^{1/2} Pr^{-1/2}$
\cite{celani2006rayleigh,boffetta2009kolmogorov,boffetta2012ultimate}
both in two and three dimensions.

\begin{figure}[h!]
\includegraphics[width=\columnwidth]{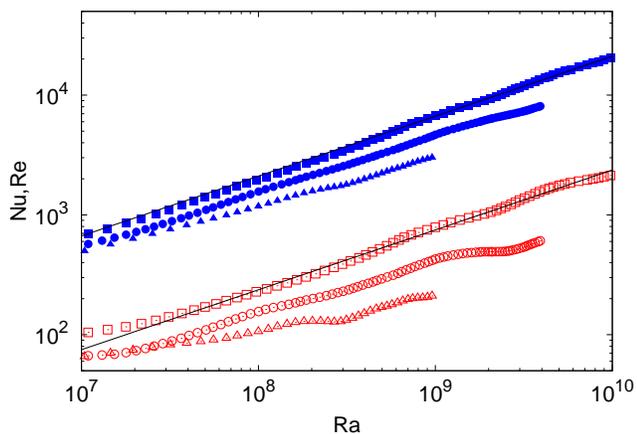}
\caption{
Reynolds number (filled blue symbols) and Nusselt number (open red symbols) 
as a function of Rayleigh number for the simulations with $\Omega \tau=0$ 
(squares),  $\Omega \tau=6.25$ (circles) and $\Omega \tau=12.5$ (triangles). 
The black lines represent the ultimate state behavior $Ra^{1/2}$.}
\label{fig3.4}
\end{figure}
We find that rotation reduces the value of $Nu$ both as a function of time 
(mainly as a consequence of the reduction of $h(t)$) and as a function 
of $Ra$, i.e. at fixed $h$ as shown in Fig.~\ref{fig3.4}.
This latter result means that also the correlation between $w$
and $T$ is reduced by rotation.
We find that also $Re$ at given $Ra$ decreases in presence of rotation, 
mainly as a consequence of the suppression of the velocity fluctuations.
Moreover we find that, at larger values of $\Omega$, the ultimate state 
scaling is violated as both $Nu$ and $Re$ show a dependence on $Ra$ 
slower than $Ra^{1/2}$. 

We remark that our results on the suppression of the heat transport 
in rotating RT turbulence are qualitatively different
to what observed in rotating boundary-forced convective flow, e.g. 
in Rayleigh-Benard (RB) convection. In the case of rotating RB convection 
both experiments \cite{liu1997heat,zhong2009prandtl,weiss2016heat} and 
numerical simulations 
\cite{kunnen2006heat,oresta2007transitional,stevens2013heat}
have shown an enhancement of $Nu$ for moderate values of rotation 
$\Omega$ with respect to the case $\Omega=0$. 
This enhancement is ascribed to Ekman pumping which extracts hot and
cold fluid from the boundary layers and contributes to the heat flux 
\cite{kunnen2006heat}. At large values of rotations (and for large
$Ra$) this mechanisms is less efficient and a reduction of $Nu$ 
is observed. Because of the absence of boundary layers in RT
convection, it is not contradictory that here we do not observe 
an increase of $Nu$ for any value of $\Omega$ and $Ra$. 

Beside the differences between RT and RB turbulence due 
to the boundary conditions, other phenomena occurring 
in the bulk of rotating convection display interesting similarities between the two systems. 
In particular, studies of rotating RB has found an increase of the 
internal mean gradient with decreasing $Ro$~\cite{julien1996rapidly,
hart1999thermal,liu2011local}
and a corresponding decrease in horizontal versus vertical fluctuations
of temperature, which indicates that rotation favors the
horizontal transport and suppresses the vertical mixing. 

\subsection{Energy spectra}
In order to investigate the degree of anisotropy at different scales, 
we computed at given times during the evolution of the system
the energy spectra of the horizontal and vertical components of the velocity, 
indicated respectively as $E_x(k)$ and $E_z(k)$. 
The spectra has been computed from the Fourier transform
of the velocity fields in the whole periodic domain, 
and are plotted as a function of the wavenumber 
$k = (k_x^2+k_y^2+k_z^2)^{1/2}$. 

\begin{figure}[h!]
\includegraphics[width=\columnwidth]{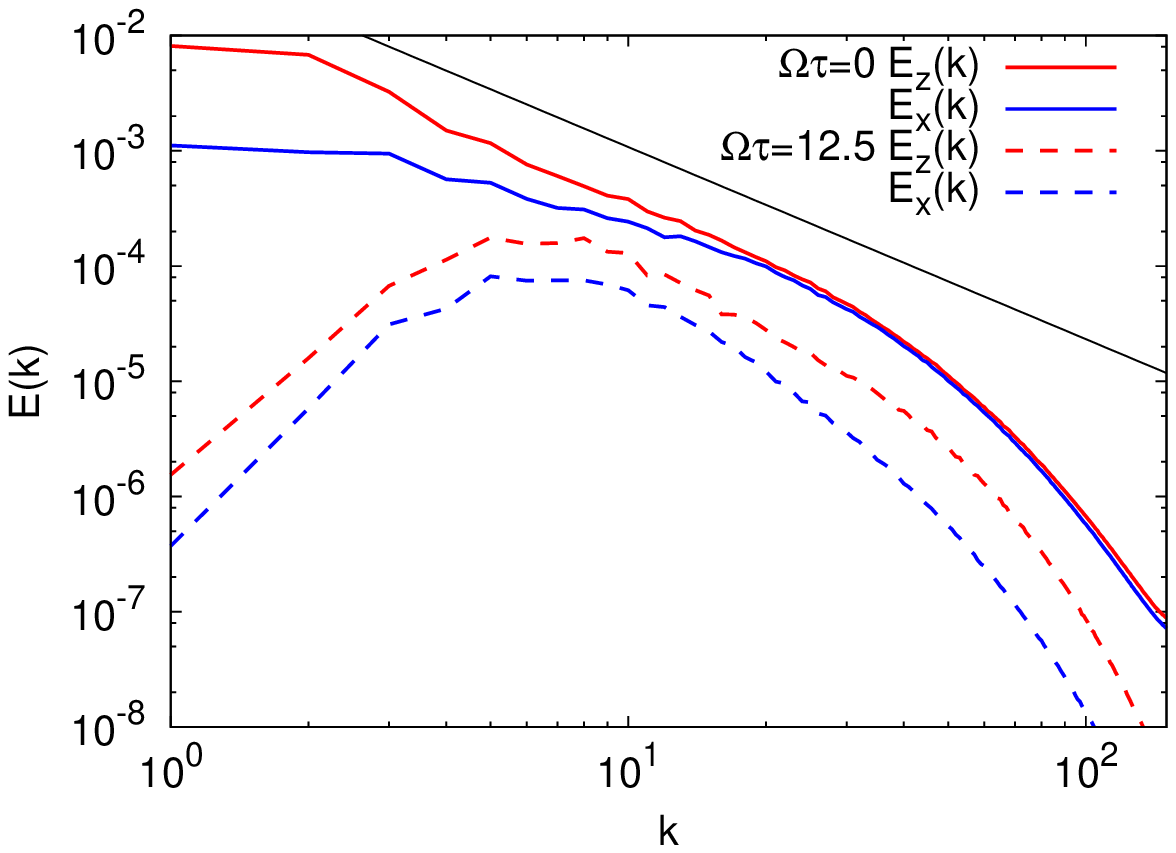}
\includegraphics[width=\columnwidth]{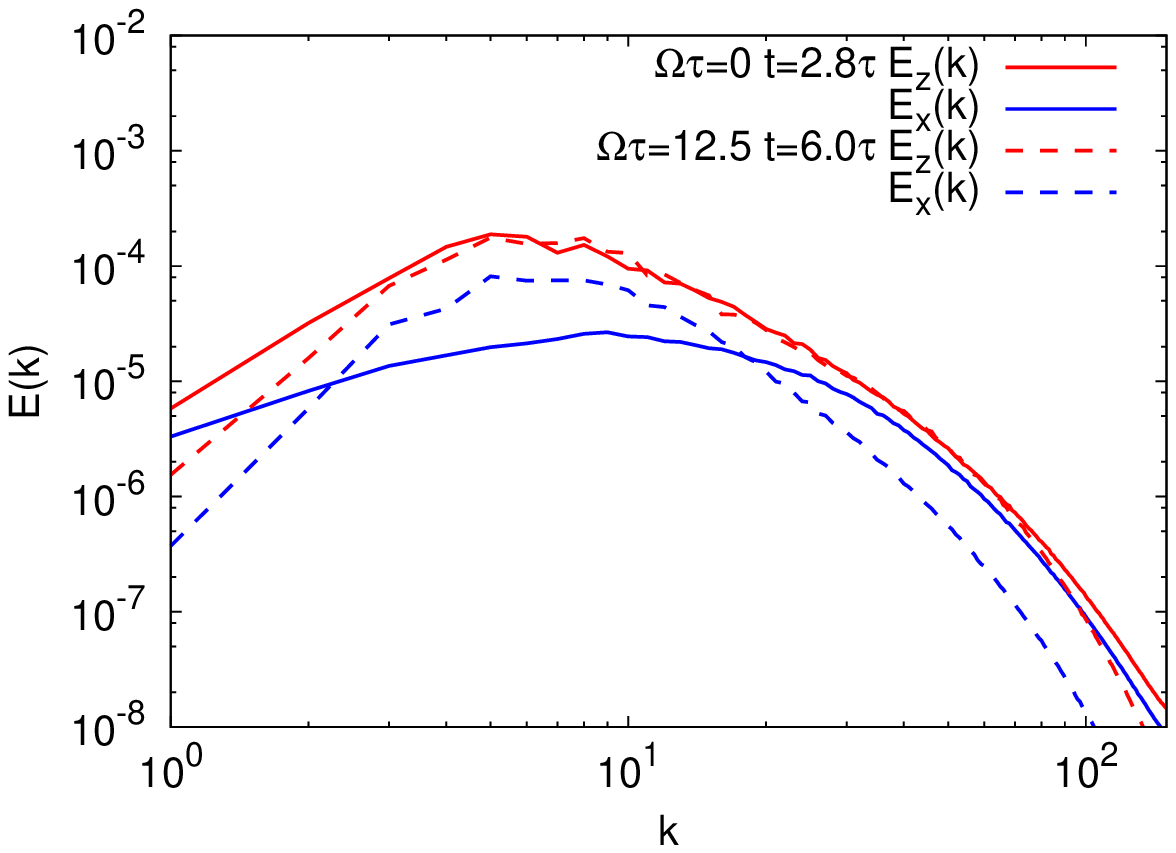}
\caption{Upper plot: Vertical $E_z(k)$ (red) and horizontal $E_x(k)$ (blue)
kinetic energy spectra computed at the time $t=6.0 \tau$ both in the
absence of rotation (continuous lines) and with $\Omega \tau=12.5$ 
(dashed lines).
Lower plot: Vertical and horizontal kinetic energy spectra computed at
time $t=2.8 \tau$ for $\Omega \tau=0$ and at time $t=6.0 \tau$ for 
$\Omega \tau=12.5$ at which the peaks of the vertical spectra are the same.}
\label{fig3.5}
\end{figure}

In Fig.~\ref{fig3.5} we compare the horizontal and vertical spectra 
computed at the same time $t=6\tau$ with and without rotation.  
We see that for $\Omega=0$ the two spectra are almost identical at 
high-wavenumbers, 
indicating the recovering of isotropy in the flow at small scales. 
\cite{boffetta2010statistics}.
The recovery of small-scale isotropy does not occur in presence of rotation.  
For $\Omega\tau = 12.5$ we find that vertical velocity
fluctuations dominates over horizontal velocity fluctuations at all scales.
These results indicate that the ratio $E_z(k)/E_x(k)$
is changed by rotation mostly at small scales, while at large scales
rotation reduces both the spectra keeping their ratio almost constant.
This is consistent with the weak dependence on $\Omega$ 
of the ratio $w_{rms}/u_{rms}$ shown in Fig.~\ref{fig3.1}. 
On the other hand, rotation changes significantly the degree of anisotropy 
of the velocity gradient. Indeed we measure that the ratio 
$(\partial_z w)_{rms}/(\partial_x u)_{rms}$ changes from a value
close to one for $\Omega=0$ \cite{boffetta2010statistics} to
about $0.65$ for $\Omega \tau=12.5$.

As rotation slows down the development of turbulence, it is interesting
to compare spectra with and without rotation at two different times at
which the peaks of the vertical energy spectrum are the same. 
The comparison of the two spectra is shown in 
Fig.~\ref{fig3.5} where we compare the case $\Omega \tau=0$ at 
$t=2.8 \tau$ with the case $\Omega \tau=12.5$ at $t=6.0 \tau$. 
Vertical spectra are almost identical (within statistical fluctuations)
while for horizontal spectra the effect of rotation is to reduce 
small-scale fluctuation in favor of large-scale components. 

\subsection{Cyclonic-anticyclonic asymmetry}
A remarkable feature of turbulent rotating flows 
is the predominance of cyclones, i.e., 
vortices co-rotating with the flow, over anticyclones. 
The cyclone-anticyclone asymmetry has been observed in the atmosphere 
\cite{hakim2005observed},  
and has been studied both in experiments and numerics 
(for a review see, e.g., \cite{godeferd2015structure}).   

Previous studies~\cite{bartello1994rotating,deusebio2014dimensional,gallet2014scale,naso2015cyclone}  
have investigated the dependence of the asymmetry on the Rossby number and the aspect ratio of the flow.
The asymmetry is found to be maximum for $Ro \simeq O(1)$, 
and vanishes when the flow is confined in a quasi two-dimensional geometry.  
These results suggest that the mechanisms which produce the asymmetry 
require the interaction between rotation and turbulent vortex stretching. 

\begin{figure}[h!]
\includegraphics[width=\columnwidth]{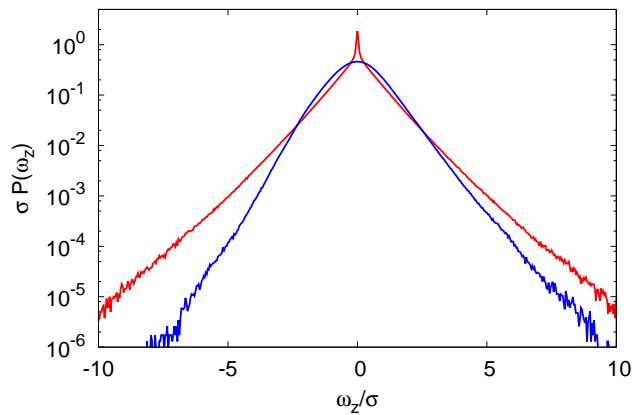} 
\caption{PDF of vertical vorticity computed in the mixing layer 
of width $h$ at time 
$t=3.2 \tau$ for $\Omega \tau=0$ (red) and $t=5.6 \tau$ for 
$\Omega \tau=12.5$ (blue). 
At these times the width of the mixing layer in the two cases is the
same.
The values of the vertical vorticity has been normalized with the standard deviation 
($\sigma = 0.010$ for $\Omega \tau=0$ and  
$\sigma = 0.039$ for  $\Omega \tau=12.5$). 
}
\label{fig3.6}
\end{figure}

Here we address the question whether or not the cyclonic-anticyclonic 
asymmetry develops also in the rotating Rayleigh-Taylor turbulence. 
To address this issue we compute the probability density function (PDFs) 
of the components of the vorticity field within the mixing layer, 
where the vorticity is significantly different from zero. 

In Figure~\ref{fig3.6} we compare the PDFs of the 
vertical component $\omega_z$ of the vorticity
with and without rotation. 
The PDFs are computed in the mixing layer of width $h$
at times at which $h$ is the same in the two cases. 
We find that in the non-rotating case the PDF is symmetric, 
while in the rotating case it is asymmetric with a positive skewness, 
which indicates the prevalence of cyclonic vortices, as
shown in Fig.~\ref{fig3.6}.
A similar result has been observed also for the rotating 
Rayleigh-B\'enard convection~\cite{julien1996rapidly}. 

A quantitative measure of this asymmetry is provided 
by the skewness of the vertical vorticity: 
\begin{equation}
S_{\omega} = {\langle \omega_z^3 \rangle \over 
\langle \omega_z^2 \rangle^{3/2}}
\label{eq:3.2}
\end{equation}
where the average is here computed within the mixing layer.

\begin{figure}[h!]
\includegraphics[width=\columnwidth]{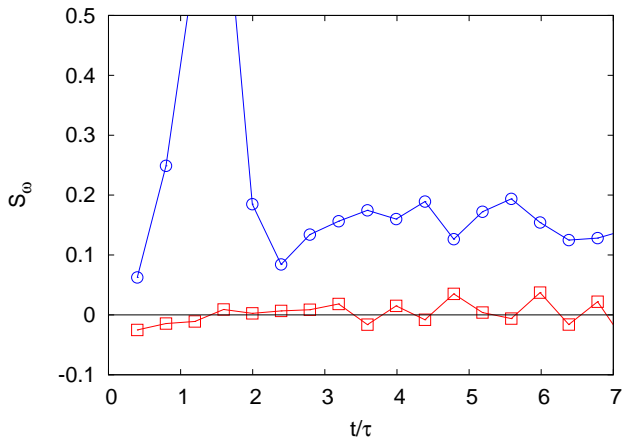}
\caption{
Time evolution of the skewness 
$S_{\omega}=\langle \omega_z^3 \rangle/\langle \omega_z^2 \rangle^{3/2}$
for $\Omega \tau=0$ (red) and $\Omega \tau=12.5$ (blue).}
\label{fig3.7}
\end{figure}

The time evolution of $S_{\omega}$ is shown in Fig.~\ref{fig3.7} for 
two cases with and without rotation. While for $\Omega=0$ we have
a skewness consistent with zero, for $\Omega \tau=12.5$ we observe, after the
development of the turbulent phase, a value $S_{\omega} \simeq 0.15$ 
almost independent on time. The weak time-dependence of the skewness is
somehow surprising, because the cyclonic-anticyclonic asymmetry is known 
to depend on the Rossby number \cite{deusebio2014dimensional}, 
which in this case decreases 
in time as $Ro \simeq 1/\Omega t$. One possible explanation is that 
is our simulations the time dependence of $Ro$ is too weak
(see Fig.~\ref{fig3.1b}) to appreciate a significant variation of the 
asymmetry.
We conclude by remarking that the other two components of the vorticity are 
symmetric also in presence of rotation.

\section{Conclusions}
\label{sec:4}
In this paper we have studied the effect of rotation along the vertical
axis on the development and the mixing properties of Rayleigh-Taylor 
turbulence in the Boussinesq approximation of incompressible flow.
We have found that rotation reduces the intensity of turbulent velocity
fluctuations, a generalization of the known stabilizing effect on the 
linear phase \cite{chandrasekhar1961hydrodynamics,carnevale2002rotational}.
As a consequence, the temporal growth of the mixing layer is observed
to be reduced with respect the case without rotation.
Rotation also reduces the efficiency of the conversion of the available 
potential energy into large-scale kinetic energy, and relatively more 
energy is dissipated at small scales.
The analysis of the kinetic energy spectra revealed that rotation induces 
a strong anisotropy at small scales, where vertical velocity fluctuations
dominate over horizontal ones. 

In spite of the reduction of the intensity of velocity fluctuations, 
we found that rotation allows them to extends in a broader region, 
a manifestation of the Taylor-Proudman theorem on the bidimensionalization
of the flow, while the scale associated to temperature fluctuations is
reduced. These results imply that in presence of rotation the mixing layer
has a complex structure which cannot be characterized by a single scale.
The presence of different characteristic scales for velocity and temperature
has important dynamical consequences. Indeed, the analysis of our numerical 
simulations show that the heat transfer, quantified by the Nusselt number,
is reduced by rotation not only as a function of time, but also at a 
given Rayleigh number (i.e. width of the mixing layer). This means
that rotation reduces the correlation $\langle w T \rangle$
between the temperature and the vertical velocity.

Finally, we have analyzed the cyclone-anticyclone asymmetry induced by 
rotation.
We have found that, as in other turbulent systems, rotation induces 
an asymmetry in the PDF of the vertical vorticity which displays a
a positive (with respect to the background rotation) skewness. 
The skewness is found to be approximatively constant during the development 
of the turbulent mixing layer.

In conclusion, we found that rotation induces a rich and complex 
phenomenology in Rayleigh-Taylor turbulence.
As RT turbulence is a prototypical example of bulk turbulent convection, 
not driven by boundary layers, we expect similar effects in other 
convective flows dominated by bulk properties. Further numerical
and experimental work in this direction would be extremely interesting.

\begin{acknowledgments}
We acknowledge support from the European COST Action No. MP1305
“Flowing Matter.”
AM thanks the financial support from the PRIN 2012 project n. D38C13000610001
funded by the Italian Ministry of Education. AM is
also grateful for the financial support for the computational infrastructure
from the Italian flagship project RITMARE.
HPC center CINECA is gratefully acknowledged for computing resources.
\end{acknowledgments}

\bibliography{biblio}

\begin{thebibliography}{52}%
\makeatletter
\providecommand \@ifxundefined [1]{%
 \@ifx{#1\undefined}
}%
\providecommand \@ifnum [1]{%
 \ifnum #1\expandafter \@firstoftwo
 \else \expandafter \@secondoftwo
 \fi
}%
\providecommand \@ifx [1]{%
 \ifx #1\expandafter \@firstoftwo
 \else \expandafter \@secondoftwo
 \fi
}%
\providecommand \natexlab [1]{#1}%
\providecommand \enquote  [1]{``#1''}%
\providecommand \bibnamefont  [1]{#1}%
\providecommand \bibfnamefont [1]{#1}%
\providecommand \citenamefont [1]{#1}%
\providecommand \href@noop [0]{\@secondoftwo}%
\providecommand \href [0]{\begingroup \@sanitize@url \@href}%
\providecommand \@href[1]{\@@startlink{#1}\@@href}%
\providecommand \@@href[1]{\endgroup#1\@@endlink}%
\providecommand \@sanitize@url [0]{\catcode `\\12\catcode `\$12\catcode
  `\&12\catcode `\#12\catcode `\^12\catcode `\_12\catcode `\%12\relax}%
\providecommand \@@startlink[1]{}%
\providecommand \@@endlink[0]{}%
\providecommand \url  [0]{\begingroup\@sanitize@url \@url }%
\providecommand \@url [1]{\endgroup\@href {#1}{\urlprefix }}%
\providecommand \urlprefix  [0]{URL }%
\providecommand \Eprint [0]{\href }%
\providecommand \doibase [0]{http://dx.doi.org/}%
\providecommand \selectlanguage [0]{\@gobble}%
\providecommand \bibinfo  [0]{\@secondoftwo}%
\providecommand \bibfield  [0]{\@secondoftwo}%
\providecommand \translation [1]{[#1]}%
\providecommand \BibitemOpen [0]{}%
\providecommand \bibitemStop [0]{}%
\providecommand \bibitemNoStop [0]{.\EOS\space}%
\providecommand \EOS [0]{\spacefactor3000\relax}%
\providecommand \BibitemShut  [1]{\csname bibitem#1\endcsname}%
\let\auto@bib@innerbib\@empty
\bibitem [{\citenamefont
  {Chandrasekhar}(1961)}]{chandrasekhar1961hydrodynamics}%
  \BibitemOpen
  \bibfield  {author} {\bibinfo {author} {\bibfnamefont {S.}~\bibnamefont
  {Chandrasekhar}},\ }\href@noop {} {\emph {\bibinfo {title} {{Hydrodynamics
  and Hydromagnetic Stability}}}}\ (\bibinfo  {publisher} {Oxford university
  press},\ \bibinfo {year} {1961})\BibitemShut {NoStop}%
\bibitem [{\citenamefont {Sharp}(1984)}]{sharp1984overview}%
  \BibitemOpen
  \bibfield  {author} {\bibinfo {author} {\bibfnamefont {D.H.}\ \bibnamefont
  {Sharp}},\ }\bibfield  {title} {\enquote {\bibinfo {title} {{An overview of
  Rayleigh-Taylor instability}},}\ }\href@noop {} {\bibfield  {journal}
  {\bibinfo  {journal} {Physica}\ }\textbf {\bibinfo {volume} {12D}},\ \bibinfo
  {pages} {3--18} (\bibinfo {year} {1984})}\BibitemShut {NoStop}%
\bibitem [{\citenamefont {Abarzhi}(2010)}]{abarzhi2010review}%
  \BibitemOpen
  \bibfield  {author} {\bibinfo {author} {\bibfnamefont {S.I.}\ \bibnamefont
  {Abarzhi}},\ }\bibfield  {title} {\enquote {\bibinfo {title} {{Review of
  theoretical modelling approaches of Rayleigh--Taylor instabilities and
  turbulent mixing}},}\ }\href@noop {} {\bibfield  {journal} {\bibinfo
  {journal} {Phil. Trans. R. Soc. Lond. A}\ }\textbf {\bibinfo {volume}
  {368}},\ \bibinfo {pages} {1809--1828} (\bibinfo {year} {2010})}\BibitemShut
  {NoStop}%
\bibitem [{\citenamefont {Boffetta}\ and\ \citenamefont
  {Mazzino}(2017)}]{boffetta2017}%
  \BibitemOpen
  \bibfield  {author} {\bibinfo {author} {\bibfnamefont {G.}~\bibnamefont
  {Boffetta}}\ and\ \bibinfo {author} {\bibfnamefont {A.}~\bibnamefont
  {Mazzino}},\ }\bibfield  {title} {\enquote {\bibinfo {title} {{Incompressible
  Rayleigh-Taylor Tubulence}},}\ }\href@noop {} {\bibfield  {journal} {\bibinfo
   {journal} {Annu. Rev. Fluid Mech.}\ }\textbf {\bibinfo {volume} {49}}
  (\bibinfo {year} {2017})}\BibitemShut {NoStop}%
\bibitem [{\citenamefont {Hillebrandt}\ and\ \citenamefont
  {Niemeyer}(2000)}]{hillebrandt2000type}%
  \BibitemOpen
  \bibfield  {author} {\bibinfo {author} {\bibfnamefont {W.}~\bibnamefont
  {Hillebrandt}}\ and\ \bibinfo {author} {\bibfnamefont {J.C.}\ \bibnamefont
  {Niemeyer}},\ }\bibfield  {title} {\enquote {\bibinfo {title} {{Type Ia
  Supernova Explosion Models}},}\ }\href@noop {} {\bibfield  {journal}
  {\bibinfo  {journal} {Annu. Rev. Astronomy Astrophys.}\ }\textbf {\bibinfo
  {volume} {38}},\ \bibinfo {pages} {191--230} (\bibinfo {year}
  {2000})}\BibitemShut {NoStop}%
\bibitem [{\citenamefont {Bell}\ \emph {et~al.}(2004)\citenamefont {Bell},
  \citenamefont {Day}, \citenamefont {Rendleman}, \citenamefont {Woosley},\
  and\ \citenamefont {Zingale}}]{bell2004direct}%
  \BibitemOpen
  \bibfield  {author} {\bibinfo {author} {\bibfnamefont {J.B.}\ \bibnamefont
  {Bell}}, \bibinfo {author} {\bibfnamefont {M.S.}\ \bibnamefont {Day}},
  \bibinfo {author} {\bibfnamefont {C.A.}\ \bibnamefont {Rendleman}}, \bibinfo
  {author} {\bibfnamefont {S.E.}\ \bibnamefont {Woosley}}, \ and\ \bibinfo
  {author} {\bibfnamefont {M.}~\bibnamefont {Zingale}},\ }\bibfield  {title}
  {\enquote {\bibinfo {title} {{Direct Numerical Simulations of Type Ia
  Supernovae Flames. II. The Rayleigh--Taylor Instability}},}\ }\href@noop {}
  {\bibfield  {journal} {\bibinfo  {journal} {Astrophys. J.}\ }\textbf
  {\bibinfo {volume} {608}},\ \bibinfo {pages} {883} (\bibinfo {year}
  {2004})}\BibitemShut {NoStop}%
\bibitem [{\citenamefont {Weinberg}\ and\ \citenamefont
  {Schmeling}(1992)}]{weinberg1992polydiapirs}%
  \BibitemOpen
  \bibfield  {author} {\bibinfo {author} {\bibfnamefont {R.F.}\ \bibnamefont
  {Weinberg}}\ and\ \bibinfo {author} {\bibfnamefont {H.}~\bibnamefont
  {Schmeling}},\ }\bibfield  {title} {\enquote {\bibinfo {title} {{Polydiapirs:
  multiwavelength gravity structures}},}\ }\href@noop {} {\bibfield  {journal}
  {\bibinfo  {journal} {J. Struct. Geol.}\ }\textbf {\bibinfo {volume} {14}},\
  \bibinfo {pages} {425--436} (\bibinfo {year} {1992})}\BibitemShut {NoStop}%
\bibitem [{\citenamefont {Agee}(1975)}]{agee1975some}%
  \BibitemOpen
  \bibfield  {author} {\bibinfo {author} {\bibfnamefont {E.M.}\ \bibnamefont
  {Agee}},\ }\bibfield  {title} {\enquote {\bibinfo {title} {{Some inferences
  of eddy viscosity associated with instabilities in the atmosphere}},}\
  }\href@noop {} {\bibfield  {journal} {\bibinfo  {journal} {J. Atmos. Sci.}\
  }\textbf {\bibinfo {volume} {32}},\ \bibinfo {pages} {642--646} (\bibinfo
  {year} {1975})}\BibitemShut {NoStop}%
\bibitem [{\citenamefont {Read}(1984)}]{read1984experimental}%
  \BibitemOpen
  \bibfield  {author} {\bibinfo {author} {\bibfnamefont {K.I.}\ \bibnamefont
  {Read}},\ }\bibfield  {title} {\enquote {\bibinfo {title} {{Experimental
  investigation of turbulent mixing by Rayleigh-Taylor instability}},}\
  }\href@noop {} {\bibfield  {journal} {\bibinfo  {journal} {Physica}\ }\textbf
  {\bibinfo {volume} {12D}},\ \bibinfo {pages} {45--58} (\bibinfo {year}
  {1984})}\BibitemShut {NoStop}%
\bibitem [{\citenamefont {Dalziel}(1993)}]{dalziel1993rayleigh}%
  \BibitemOpen
  \bibfield  {author} {\bibinfo {author} {\bibfnamefont {S.B.}\ \bibnamefont
  {Dalziel}},\ }\bibfield  {title} {\enquote {\bibinfo {title}
  {{Rayleigh-Taylor instability: experiments with image analysis}},}\
  }\href@noop {} {\bibfield  {journal} {\bibinfo  {journal} {Dyn. Atmos.
  Oceans}\ }\textbf {\bibinfo {volume} {20}},\ \bibinfo {pages} {127--153}
  (\bibinfo {year} {1993})}\BibitemShut {NoStop}%
\bibitem [{\citenamefont {Dimonte}\ and\ \citenamefont
  {Schneider}(1996)}]{dimonte1996turbulent}%
  \BibitemOpen
  \bibfield  {author} {\bibinfo {author} {\bibfnamefont {G.}~\bibnamefont
  {Dimonte}}\ and\ \bibinfo {author} {\bibfnamefont {M.B.}\ \bibnamefont
  {Schneider}},\ }\bibfield  {title} {\enquote {\bibinfo {title} {{Turbulent
  Rayleigh-Taylor instability experiments with variable acceleration}},}\
  }\href@noop {} {\bibfield  {journal} {\bibinfo  {journal} {Phys. Rev. E}\
  }\textbf {\bibinfo {volume} {54}},\ \bibinfo {pages} {3740} (\bibinfo {year}
  {1996})}\BibitemShut {NoStop}%
\bibitem [{\citenamefont {Banerjee}\ and\ \citenamefont
  {Andrews}(2006)}]{banerjee2006statistically}%
  \BibitemOpen
  \bibfield  {author} {\bibinfo {author} {\bibfnamefont {A.}~\bibnamefont
  {Banerjee}}\ and\ \bibinfo {author} {\bibfnamefont {M.J.}\ \bibnamefont
  {Andrews}},\ }\bibfield  {title} {\enquote {\bibinfo {title} {{Statistically
  steady measurements of Rayleigh-Taylor mixing in a gas channel}},}\
  }\href@noop {} {\bibfield  {journal} {\bibinfo  {journal} {Phys. Fluids}\
  }\textbf {\bibinfo {volume} {18}},\ \bibinfo {pages} {035107} (\bibinfo
  {year} {2006})}\BibitemShut {NoStop}%
\bibitem [{\citenamefont {Andrews}\ and\ \citenamefont
  {Dalziel}(2010)}]{andrews2010small}%
  \BibitemOpen
  \bibfield  {author} {\bibinfo {author} {\bibfnamefont {M.J.}\ \bibnamefont
  {Andrews}}\ and\ \bibinfo {author} {\bibfnamefont {S.B.}\ \bibnamefont
  {Dalziel}},\ }\bibfield  {title} {\enquote {\bibinfo {title} {{Small Atwood
  number Rayleigh--Taylor experiments}},}\ }\href@noop {} {\bibfield  {journal}
  {\bibinfo  {journal} {Phil. Trans. R. Soc. A}\ }\textbf {\bibinfo {volume}
  {368}},\ \bibinfo {pages} {1663--1679} (\bibinfo {year} {2010})}\BibitemShut
  {NoStop}%
\bibitem [{\citenamefont {Banerjee}\ \emph {et~al.}(2010)\citenamefont
  {Banerjee}, \citenamefont {Kraft},\ and\ \citenamefont
  {Andrews}}]{banerjee2010detailed}%
  \BibitemOpen
  \bibfield  {author} {\bibinfo {author} {\bibfnamefont {A.}~\bibnamefont
  {Banerjee}}, \bibinfo {author} {\bibfnamefont {W.N.}\ \bibnamefont {Kraft}},
  \ and\ \bibinfo {author} {\bibfnamefont {M.J.}\ \bibnamefont {Andrews}},\
  }\bibfield  {title} {\enquote {\bibinfo {title} {{Detailed measurements of a
  statistically steady Rayleigh--Taylor mixing layer from small to high Atwood
  numbers}},}\ }\href@noop {} {\bibfield  {journal} {\bibinfo  {journal} {J.
  Fluid Mech.}\ }\textbf {\bibinfo {volume} {659}},\ \bibinfo {pages}
  {127--190} (\bibinfo {year} {2010})}\BibitemShut {NoStop}%
\bibitem [{\citenamefont {Young}\ \emph {et~al.}(2001)\citenamefont {Young},
  \citenamefont {Tufo}, \citenamefont {Dubey},\ and\ \citenamefont
  {Rosner}}]{young2001on}%
  \BibitemOpen
  \bibfield  {author} {\bibinfo {author} {\bibfnamefont {Y.N.}\ \bibnamefont
  {Young}}, \bibinfo {author} {\bibfnamefont {H.}~\bibnamefont {Tufo}},
  \bibinfo {author} {\bibfnamefont {A.}~\bibnamefont {Dubey}}, \ and\ \bibinfo
  {author} {\bibfnamefont {R.}~\bibnamefont {Rosner}},\ }\bibfield  {title}
  {\enquote {\bibinfo {title} {{On the miscible Rayleigh--Taylor instability:
  two and three dimensions}},}\ }\href@noop {} {\bibfield  {journal} {\bibinfo
  {journal} {J. Fluid Mech.}\ }\textbf {\bibinfo {volume} {447}},\ \bibinfo
  {pages} {377--408} (\bibinfo {year} {2001})}\BibitemShut {NoStop}%
\bibitem [{\citenamefont {Dimonte}\ \emph {et~al.}(2004)\citenamefont
  {Dimonte}, \citenamefont {Youngs}, \citenamefont {Dimits}, \citenamefont
  {Weber}, \citenamefont {Marinak}, \citenamefont {Wunsch}, \citenamefont
  {Garasi}, \citenamefont {Robinson}, \citenamefont {Andrews}, \citenamefont
  {Ramaprabhu} \emph {et~al.}}]{dimonte2004comparative}%
  \BibitemOpen
  \bibfield  {author} {\bibinfo {author} {\bibfnamefont {G.}~\bibnamefont
  {Dimonte}}, \bibinfo {author} {\bibfnamefont {D.L.}\ \bibnamefont {Youngs}},
  \bibinfo {author} {\bibfnamefont {A.}~\bibnamefont {Dimits}}, \bibinfo
  {author} {\bibfnamefont {S.}~\bibnamefont {Weber}}, \bibinfo {author}
  {\bibfnamefont {M.}~\bibnamefont {Marinak}}, \bibinfo {author} {\bibfnamefont
  {S.}~\bibnamefont {Wunsch}}, \bibinfo {author} {\bibfnamefont
  {C.}~\bibnamefont {Garasi}}, \bibinfo {author} {\bibfnamefont
  {A.}~\bibnamefont {Robinson}}, \bibinfo {author} {\bibfnamefont {M.J.}\
  \bibnamefont {Andrews}}, \bibinfo {author} {\bibfnamefont {P.}~\bibnamefont
  {Ramaprabhu}},  \emph {et~al.},\ }\bibfield  {title} {\enquote {\bibinfo
  {title} {{A comparative study of the turbulent Rayleigh--Taylor instability
  using high-resolution three-dimensional numerical simulations: the
  Alpha-Group collaboration}},}\ }\href@noop {} {\bibfield  {journal} {\bibinfo
   {journal} {Phys. Fluids}\ }\textbf {\bibinfo {volume} {16}},\ \bibinfo
  {pages} {1668--1693} (\bibinfo {year} {2004})}\BibitemShut {NoStop}%
\bibitem [{\citenamefont {Cabot}\ and\ \citenamefont
  {Cook}(2006)}]{cabot2006reynolds}%
  \BibitemOpen
  \bibfield  {author} {\bibinfo {author} {\bibfnamefont {W.H.}\ \bibnamefont
  {Cabot}}\ and\ \bibinfo {author} {\bibfnamefont {A.W.}\ \bibnamefont
  {Cook}},\ }\bibfield  {title} {\enquote {\bibinfo {title} {{Reynolds number
  effects on Rayleigh--Taylor instability with possible implications for type
  Ia supernovae}},}\ }\href@noop {} {\bibfield  {journal} {\bibinfo  {journal}
  {Nat. Phys.}\ ,\ \bibinfo {pages} {562}} (\bibinfo {year}
  {2006})}\BibitemShut {NoStop}%
\bibitem [{\citenamefont {Vladimirova}\ and\ \citenamefont
  {Chertkov}(2009)}]{vladimirova2009self}%
  \BibitemOpen
  \bibfield  {author} {\bibinfo {author} {\bibfnamefont {N.}~\bibnamefont
  {Vladimirova}}\ and\ \bibinfo {author} {\bibfnamefont {M.}~\bibnamefont
  {Chertkov}},\ }\bibfield  {title} {\enquote {\bibinfo {title}
  {{Self-similarity and universality in Rayleigh--Taylor, Boussinesq
  turbulence}},}\ }\href@noop {} {\bibfield  {journal} {\bibinfo  {journal}
  {Phys. Fluids}\ }\textbf {\bibinfo {volume} {21}},\ \bibinfo {pages} {015102}
  (\bibinfo {year} {2009})}\BibitemShut {NoStop}%
\bibitem [{\citenamefont {Matsumoto}(2009)}]{matsumoto2009anomalous}%
  \BibitemOpen
  \bibfield  {author} {\bibinfo {author} {\bibfnamefont {T.}~\bibnamefont
  {Matsumoto}},\ }\bibfield  {title} {\enquote {\bibinfo {title} {{Anomalous
  scaling of three-dimensional Rayleigh-Taylor turbulence}},}\ }\href@noop {}
  {\bibfield  {journal} {\bibinfo  {journal} {Phys. Rev. E}\ }\textbf {\bibinfo
  {volume} {79}},\ \bibinfo {pages} {055301} (\bibinfo {year}
  {2009})}\BibitemShut {NoStop}%
\bibitem [{\citenamefont {Biferale}\ \emph {et~al.}(2010)\citenamefont
  {Biferale}, \citenamefont {Mantovani}, \citenamefont {Sbragaglia},
  \citenamefont {Scagliarini}, \citenamefont {Toschi},\ and\ \citenamefont
  {Tripiccione}}]{biferale2010high}%
  \BibitemOpen
  \bibfield  {author} {\bibinfo {author} {\bibfnamefont {L.}~\bibnamefont
  {Biferale}}, \bibinfo {author} {\bibfnamefont {F.}~\bibnamefont {Mantovani}},
  \bibinfo {author} {\bibfnamefont {M.}~\bibnamefont {Sbragaglia}}, \bibinfo
  {author} {\bibfnamefont {A.}~\bibnamefont {Scagliarini}}, \bibinfo {author}
  {\bibfnamefont {F.}~\bibnamefont {Toschi}}, \ and\ \bibinfo {author}
  {\bibfnamefont {R.}~\bibnamefont {Tripiccione}},\ }\bibfield  {title}
  {\enquote {\bibinfo {title} {{High resolution numerical study of
  Rayleigh--Taylor turbulence using a thermal lattice Boltzmann scheme}},}\
  }\href@noop {} {\bibfield  {journal} {\bibinfo  {journal} {Phys. Fluids}\
  }\textbf {\bibinfo {volume} {22}},\ \bibinfo {pages} {115112} (\bibinfo
  {year} {2010})}\BibitemShut {NoStop}%
\bibitem [{\citenamefont {Chertkov}(2003)}]{chertkov2003phenomenology}%
  \BibitemOpen
  \bibfield  {author} {\bibinfo {author} {\bibfnamefont {M.}~\bibnamefont
  {Chertkov}},\ }\bibfield  {title} {\enquote {\bibinfo {title} {{Phenomenology
  of Rayleigh-Taylor Turbulence}},}\ }\href@noop {} {\bibfield  {journal}
  {\bibinfo  {journal} {Phys. Rev. Lett.}\ }\textbf {\bibinfo {volume} {91}},\
  \bibinfo {pages} {115001} (\bibinfo {year} {2003})}\BibitemShut {NoStop}%
\bibitem [{\citenamefont {Mitchner}\ and\ \citenamefont
  {Landshoff}(1964)}]{mitchner1964rayleigh}%
  \BibitemOpen
  \bibfield  {author} {\bibinfo {author} {\bibfnamefont {M.}~\bibnamefont
  {Mitchner}}\ and\ \bibinfo {author} {\bibfnamefont {R.K.M.}\ \bibnamefont
  {Landshoff}},\ }\bibfield  {title} {\enquote {\bibinfo {title}
  {{Rayleigh-Taylor Instability for Compressible Fluids}},}\ }\href@noop {}
  {\bibfield  {journal} {\bibinfo  {journal} {Phys. Fluids}\ }\textbf {\bibinfo
  {volume} {7}},\ \bibinfo {pages} {862--866} (\bibinfo {year}
  {1964})}\BibitemShut {NoStop}%
\bibitem [{\citenamefont {Scagliarini}\ \emph {et~al.}(2010)\citenamefont
  {Scagliarini}, \citenamefont {Biferale}, \citenamefont {Sbragaglia},
  \citenamefont {Sugiyama},\ and\ \citenamefont
  {Toschi}}]{scagliarini2010lattice}%
  \BibitemOpen
  \bibfield  {author} {\bibinfo {author} {\bibfnamefont {A.}~\bibnamefont
  {Scagliarini}}, \bibinfo {author} {\bibfnamefont {L.}~\bibnamefont
  {Biferale}}, \bibinfo {author} {\bibfnamefont {M.}~\bibnamefont
  {Sbragaglia}}, \bibinfo {author} {\bibfnamefont {K.}~\bibnamefont
  {Sugiyama}}, \ and\ \bibinfo {author} {\bibfnamefont {F.}~\bibnamefont
  {Toschi}},\ }\bibfield  {title} {\enquote {\bibinfo {title} {{Lattice
  Boltzmann methods for thermal flows: Continuum limit and applications to
  compressible Rayleigh--Taylor systems}},}\ }\href@noop {} {\bibfield
  {journal} {\bibinfo  {journal} {Phys. Fluids}\ }\textbf {\bibinfo {volume}
  {22}},\ \bibinfo {pages} {055101} (\bibinfo {year} {2010})}\BibitemShut
  {NoStop}%
\bibitem [{\citenamefont {Boffetta}\ \emph
  {et~al.}(2010{\natexlab{a}})\citenamefont {Boffetta}, \citenamefont
  {Mazzino}, \citenamefont {Musacchio},\ and\ \citenamefont
  {Vozella}}]{boffetta2010polymer}%
  \BibitemOpen
  \bibfield  {author} {\bibinfo {author} {\bibfnamefont {G.}~\bibnamefont
  {Boffetta}}, \bibinfo {author} {\bibfnamefont {A.}~\bibnamefont {Mazzino}},
  \bibinfo {author} {\bibfnamefont {S.}~\bibnamefont {Musacchio}}, \ and\
  \bibinfo {author} {\bibfnamefont {L.}~\bibnamefont {Vozella}},\ }\bibfield
  {title} {\enquote {\bibinfo {title} {{Polymer heat transport enhancement in
  thermal convection: The case of Rayleigh-Taylor turbulence}},}\ }\href@noop
  {} {\bibfield  {journal} {\bibinfo  {journal} {Phys. Rev. Lett.}\ }\textbf
  {\bibinfo {volume} {104}},\ \bibinfo {pages} {184501} (\bibinfo {year}
  {2010}{\natexlab{a}})}\BibitemShut {NoStop}%
\bibitem [{\citenamefont {Chertkov}\ \emph {et~al.}(2005)\citenamefont
  {Chertkov}, \citenamefont {Kolokolov},\ and\ \citenamefont
  {Lebedev}}]{chertkov2005effects}%
  \BibitemOpen
  \bibfield  {author} {\bibinfo {author} {\bibfnamefont {M.}~\bibnamefont
  {Chertkov}}, \bibinfo {author} {\bibfnamefont {I.}~\bibnamefont {Kolokolov}},
  \ and\ \bibinfo {author} {\bibfnamefont {V.}~\bibnamefont {Lebedev}},\
  }\bibfield  {title} {\enquote {\bibinfo {title} {{Effects of surface tension
  on immiscible Rayleigh-Taylor turbulence}},}\ }\href@noop {} {\bibfield
  {journal} {\bibinfo  {journal} {Phys. Rev. E}\ }\textbf {\bibinfo {volume}
  {71}},\ \bibinfo {pages} {055301} (\bibinfo {year} {2005})}\BibitemShut
  {NoStop}%
\bibitem [{\citenamefont {Lawrie}\ and\ \citenamefont
  {Dalziel}(2011)}]{lawrie2011turbulent}%
  \BibitemOpen
  \bibfield  {author} {\bibinfo {author} {\bibfnamefont {A.G.W.}\ \bibnamefont
  {Lawrie}}\ and\ \bibinfo {author} {\bibfnamefont {S.B.}\ \bibnamefont
  {Dalziel}},\ }\bibfield  {title} {\enquote {\bibinfo {title} {{Turbulent
  diffusion in tall tubes. I. Models for Rayleigh-Taylor instability}},}\
  }\href@noop {} {\bibfield  {journal} {\bibinfo  {journal} {Phys. Fluids}\
  }\textbf {\bibinfo {volume} {23}},\ \bibinfo {pages} {085109} (\bibinfo
  {year} {2011})}\BibitemShut {NoStop}%
\bibitem [{\citenamefont {Boffetta}\ \emph
  {et~al.}(2012{\natexlab{a}})\citenamefont {Boffetta}, \citenamefont
  {De~Lillo}, \citenamefont {Mazzino},\ and\ \citenamefont
  {Musacchio}}]{boffetta2012bolgiano}%
  \BibitemOpen
  \bibfield  {author} {\bibinfo {author} {\bibfnamefont {G.}~\bibnamefont
  {Boffetta}}, \bibinfo {author} {\bibfnamefont {F.}~\bibnamefont {De~Lillo}},
  \bibinfo {author} {\bibfnamefont {A.}~\bibnamefont {Mazzino}}, \ and\
  \bibinfo {author} {\bibfnamefont {S.}~\bibnamefont {Musacchio}},\ }\bibfield
  {title} {\enquote {\bibinfo {title} {{Bolgiano scale in confined
  Rayleigh--Taylor turbulence}},}\ }\href@noop {} {\bibfield  {journal}
  {\bibinfo  {journal} {J. Fluid Mech.}\ }\textbf {\bibinfo {volume} {690}},\
  \bibinfo {pages} {426--440} (\bibinfo {year}
  {2012}{\natexlab{a}})}\BibitemShut {NoStop}%
\bibitem [{\citenamefont {Tao}\ \emph {et~al.}(2013)\citenamefont {Tao},
  \citenamefont {He}, \citenamefont {Ye},\ and\ \citenamefont
  {Busse}}]{tao2013nonlinear}%
  \BibitemOpen
  \bibfield  {author} {\bibinfo {author} {\bibfnamefont {J.J.}\ \bibnamefont
  {Tao}}, \bibinfo {author} {\bibfnamefont {X.T.}\ \bibnamefont {He}}, \bibinfo
  {author} {\bibfnamefont {W.H.}\ \bibnamefont {Ye}}, \ and\ \bibinfo {author}
  {\bibfnamefont {F.H.}\ \bibnamefont {Busse}},\ }\bibfield  {title} {\enquote
  {\bibinfo {title} {{Nonlinear Rayleigh-Taylor instability of rotating
  inviscid fluids}},}\ }\href@noop {} {\bibfield  {journal} {\bibinfo
  {journal} {Phys. Rev. E}\ }\textbf {\bibinfo {volume} {87}},\ \bibinfo
  {pages} {013001} (\bibinfo {year} {2013})}\BibitemShut {NoStop}%
\bibitem [{\citenamefont {Carnevale}\ \emph {et~al.}(2002)\citenamefont
  {Carnevale}, \citenamefont {Orlandi}, \citenamefont {Zhou},\ and\
  \citenamefont {Kloosterziel}}]{carnevale2002rotational}%
  \BibitemOpen
  \bibfield  {author} {\bibinfo {author} {\bibfnamefont {G.F.}\ \bibnamefont
  {Carnevale}}, \bibinfo {author} {\bibfnamefont {P.}~\bibnamefont {Orlandi}},
  \bibinfo {author} {\bibfnamefont {Y.}~\bibnamefont {Zhou}}, \ and\ \bibinfo
  {author} {\bibfnamefont {R.C.}\ \bibnamefont {Kloosterziel}},\ }\bibfield
  {title} {\enquote {\bibinfo {title} {{Rotational suppression of
  Rayleigh--Taylor instability}},}\ }\href@noop {} {\bibfield  {journal}
  {\bibinfo  {journal} {J. Fluid Mech.}\ }\textbf {\bibinfo {volume} {457}},\
  \bibinfo {pages} {181--190} (\bibinfo {year} {2002})}\BibitemShut {NoStop}%
\bibitem [{\citenamefont {Baldwin}\ \emph {et~al.}(2015)\citenamefont
  {Baldwin}, \citenamefont {Scase},\ and\ \citenamefont
  {Hill}}]{baldwin2015inhibition}%
  \BibitemOpen
  \bibfield  {author} {\bibinfo {author} {\bibfnamefont {K.A.}\ \bibnamefont
  {Baldwin}}, \bibinfo {author} {\bibfnamefont {M.M.}\ \bibnamefont {Scase}}, \
  and\ \bibinfo {author} {\bibfnamefont {R.J.A.}\ \bibnamefont {Hill}},\
  }\bibfield  {title} {\enquote {\bibinfo {title} {{The Inhibition of the
  Rayleigh--Taylor Instability by Rotation}},}\ }\href@noop {} {\bibfield
  {journal} {\bibinfo  {journal} {Sci. Rep.}\ }\textbf {\bibinfo {volume}
  {5}},\ \bibinfo {pages} {11706} (\bibinfo {year} {2015})}\BibitemShut
  {NoStop}%
\bibitem [{\citenamefont {Boffetta}\ \emph
  {et~al.}(2010{\natexlab{b}})\citenamefont {Boffetta}, \citenamefont
  {Mazzino}, \citenamefont {Musacchio},\ and\ \citenamefont
  {Vozella}}]{boffetta2010statistics}%
  \BibitemOpen
  \bibfield  {author} {\bibinfo {author} {\bibfnamefont {G.}~\bibnamefont
  {Boffetta}}, \bibinfo {author} {\bibfnamefont {A.}~\bibnamefont {Mazzino}},
  \bibinfo {author} {\bibfnamefont {S.}~\bibnamefont {Musacchio}}, \ and\
  \bibinfo {author} {\bibfnamefont {L.}~\bibnamefont {Vozella}},\ }\bibfield
  {title} {\enquote {\bibinfo {title} {{Statistics of mixing in
  three-dimensional Rayleigh--Taylor turbulence at low Atwood number and
  Prandtl number one}},}\ }\href@noop {} {\bibfield  {journal} {\bibinfo
  {journal} {Phys. Fluids}\ }\textbf {\bibinfo {volume} {22}},\ \bibinfo
  {pages} {035109} (\bibinfo {year} {2010}{\natexlab{b}})}\BibitemShut
  {NoStop}%
\bibitem [{\citenamefont {Dalziel}\ \emph {et~al.}(1999)\citenamefont
  {Dalziel}, \citenamefont {Linden},\ and\ \citenamefont
  {Youngs}}]{dalziel1999self}%
  \BibitemOpen
  \bibfield  {author} {\bibinfo {author} {\bibfnamefont {S.B.}\ \bibnamefont
  {Dalziel}}, \bibinfo {author} {\bibfnamefont {P.F.}\ \bibnamefont {Linden}},
  \ and\ \bibinfo {author} {\bibfnamefont {D.L.}\ \bibnamefont {Youngs}},\
  }\bibfield  {title} {\enquote {\bibinfo {title} {{Self-similarity and
  internal structure of turbulence induced by Rayleigh--Taylor instability}},}\
  }\href@noop {} {\bibfield  {journal} {\bibinfo  {journal} {J. Fluid Mech.}\
  }\textbf {\bibinfo {volume} {399}},\ \bibinfo {pages} {1--48} (\bibinfo
  {year} {1999})}\BibitemShut {NoStop}%
\bibitem [{\citenamefont {Boffetta}\ \emph
  {et~al.}(2010{\natexlab{c}})\citenamefont {Boffetta}, \citenamefont
  {De~Lillo},\ and\ \citenamefont {Musacchio}}]{boffetta2010nonlinear}%
  \BibitemOpen
  \bibfield  {author} {\bibinfo {author} {\bibfnamefont {G.}~\bibnamefont
  {Boffetta}}, \bibinfo {author} {\bibfnamefont {F.}~\bibnamefont {De~Lillo}},
  \ and\ \bibinfo {author} {\bibfnamefont {S.}~\bibnamefont {Musacchio}},\
  }\bibfield  {title} {\enquote {\bibinfo {title} {{Nonlinear diffusion model
  for Rayleigh-Taylor mixing}},}\ }\href@noop {} {\bibfield  {journal}
  {\bibinfo  {journal} {Phys. Rev. Lett.}\ }\textbf {\bibinfo {volume} {104}},\
  \bibinfo {pages} {034505} (\bibinfo {year} {2010}{\natexlab{c}})}\BibitemShut
  {NoStop}%
\bibitem [{\citenamefont {Tritton}(1988)}]{tritton1988physical}%
  \BibitemOpen
  \bibfield  {author} {\bibinfo {author} {\bibfnamefont {D.J.}\ \bibnamefont
  {Tritton}},\ }\href@noop {} {\emph {\bibinfo {title} {{Physical fluid
  dynamics}}}}\ (\bibinfo  {publisher} {Oxford Clarendon Press},\ \bibinfo
  {year} {1988})\BibitemShut {NoStop}%
\bibitem [{\citenamefont {Celani}\ \emph {et~al.}(2006)\citenamefont {Celani},
  \citenamefont {Mazzino},\ and\ \citenamefont {Vozella}}]{celani2006rayleigh}%
  \BibitemOpen
  \bibfield  {author} {\bibinfo {author} {\bibfnamefont {A.}~\bibnamefont
  {Celani}}, \bibinfo {author} {\bibfnamefont {A.}~\bibnamefont {Mazzino}}, \
  and\ \bibinfo {author} {\bibfnamefont {L.}~\bibnamefont {Vozella}},\
  }\bibfield  {title} {\enquote {\bibinfo {title} {{Rayleigh--Taylor Turbulence
  in Two Dimensions}},}\ }\href@noop {} {\bibfield  {journal} {\bibinfo
  {journal} {Phys. Rev. Lett.}\ }\textbf {\bibinfo {volume} {96}},\ \bibinfo
  {pages} {134504} (\bibinfo {year} {2006})}\BibitemShut {NoStop}%
\bibitem [{\citenamefont {Boffetta}\ \emph {et~al.}(2009)\citenamefont
  {Boffetta}, \citenamefont {Mazzino}, \citenamefont {Musacchio},\ and\
  \citenamefont {Vozella}}]{boffetta2009kolmogorov}%
  \BibitemOpen
  \bibfield  {author} {\bibinfo {author} {\bibfnamefont {G.}~\bibnamefont
  {Boffetta}}, \bibinfo {author} {\bibfnamefont {A.}~\bibnamefont {Mazzino}},
  \bibinfo {author} {\bibfnamefont {S.}~\bibnamefont {Musacchio}}, \ and\
  \bibinfo {author} {\bibfnamefont {L.}~\bibnamefont {Vozella}},\ }\bibfield
  {title} {\enquote {\bibinfo {title} {{Kolmogorov scaling and intermittency in
  Rayleigh--Taylor turbulence}},}\ }\href@noop {} {\bibfield  {journal}
  {\bibinfo  {journal} {Phys. Rev. E}\ }\textbf {\bibinfo {volume} {79}},\
  \bibinfo {pages} {065301} (\bibinfo {year} {2009})}\BibitemShut {NoStop}%
\bibitem [{\citenamefont {Boffetta}\ \emph
  {et~al.}(2012{\natexlab{b}})\citenamefont {Boffetta}, \citenamefont
  {De~Lillo}, \citenamefont {Mazzino},\ and\ \citenamefont
  {Vozella}}]{boffetta2012ultimate}%
  \BibitemOpen
  \bibfield  {author} {\bibinfo {author} {\bibfnamefont {G.}~\bibnamefont
  {Boffetta}}, \bibinfo {author} {\bibfnamefont {F.}~\bibnamefont {De~Lillo}},
  \bibinfo {author} {\bibfnamefont {A.}~\bibnamefont {Mazzino}}, \ and\
  \bibinfo {author} {\bibfnamefont {L.}~\bibnamefont {Vozella}},\ }\bibfield
  {title} {\enquote {\bibinfo {title} {{The ultimate state of thermal
  convection in Rayleigh–Taylor turbulence}},}\ }\href@noop {} {\bibfield
  {journal} {\bibinfo  {journal} {Physica}\ }\textbf {\bibinfo {volume}
  {241D}},\ \bibinfo {pages} {137--140} (\bibinfo {year}
  {2012}{\natexlab{b}})}\BibitemShut {NoStop}%
\bibitem [{\citenamefont {Liu}\ and\ \citenamefont {Ecke}(1997)}]{liu1997heat}%
  \BibitemOpen
  \bibfield  {author} {\bibinfo {author} {\bibfnamefont {Y.}~\bibnamefont
  {Liu}}\ and\ \bibinfo {author} {\bibfnamefont {R.E.}\ \bibnamefont {Ecke}},\
  }\bibfield  {title} {\enquote {\bibinfo {title} {Heat transport scaling in
  turbulent rayleigh-b\'enard convection: Effects of rotation and prandtl
  number},}\ }\href@noop {} {\bibfield  {journal} {\bibinfo  {journal} {Phys.
  Rev. Lett.}\ }\textbf {\bibinfo {volume} {79}},\ \bibinfo {pages}
  {2257--2260} (\bibinfo {year} {1997})}\BibitemShut {NoStop}%
\bibitem [{\citenamefont {Zhong}\ \emph {et~al.}(2009)\citenamefont {Zhong},
  \citenamefont {Stevens}, \citenamefont {Clercx}, \citenamefont {Verzicco},
  \citenamefont {Lohse},\ and\ \citenamefont {Ahlers}}]{zhong2009prandtl}%
  \BibitemOpen
  \bibfield  {author} {\bibinfo {author} {\bibfnamefont {J.Q.}\ \bibnamefont
  {Zhong}}, \bibinfo {author} {\bibfnamefont {R.J.A.M.}\ \bibnamefont
  {Stevens}}, \bibinfo {author} {\bibfnamefont {H.J.H.}\ \bibnamefont
  {Clercx}}, \bibinfo {author} {\bibfnamefont {R.}~\bibnamefont {Verzicco}},
  \bibinfo {author} {\bibfnamefont {D.}~\bibnamefont {Lohse}}, \ and\ \bibinfo
  {author} {\bibfnamefont {G.}~\bibnamefont {Ahlers}},\ }\bibfield  {title}
  {\enquote {\bibinfo {title} {Prandtl-, rayleigh-, and rossby-number
  dependence of heat transport in turbulent rotating rayleigh-b\'enard
  convection},}\ }\href@noop {} {\bibfield  {journal} {\bibinfo  {journal}
  {Phys. Rev. Lett.}\ }\textbf {\bibinfo {volume} {102}},\ \bibinfo {pages}
  {044502} (\bibinfo {year} {2009})}\BibitemShut {NoStop}%
\bibitem [{\citenamefont {Weiss}\ \emph {et~al.}(2016)\citenamefont {Weiss},
  \citenamefont {Wei},\ and\ \citenamefont {Ahlers}}]{weiss2016heat}%
  \BibitemOpen
  \bibfield  {author} {\bibinfo {author} {\bibfnamefont {S.}~\bibnamefont
  {Weiss}}, \bibinfo {author} {\bibfnamefont {P.}~\bibnamefont {Wei}}, \ and\
  \bibinfo {author} {\bibfnamefont {G.}~\bibnamefont {Ahlers}},\ }\bibfield
  {title} {\enquote {\bibinfo {title} {Heat-transport enhancement in rotating
  turbulent rayleigh-b\'enard convection},}\ }\href@noop {} {\bibfield
  {journal} {\bibinfo  {journal} {Phys. Rev. E}\ }\textbf {\bibinfo {volume}
  {93}},\ \bibinfo {pages} {043102} (\bibinfo {year} {2016})}\BibitemShut
  {NoStop}%
\bibitem [{\citenamefont {Kunnen}\ \emph {et~al.}(2006)\citenamefont {Kunnen},
  \citenamefont {Clercx},\ and\ \citenamefont {Geurts}}]{kunnen2006heat}%
  \BibitemOpen
  \bibfield  {author} {\bibinfo {author} {\bibfnamefont {R.P.J.}\ \bibnamefont
  {Kunnen}}, \bibinfo {author} {\bibfnamefont {H.J.H.}\ \bibnamefont {Clercx}},
  \ and\ \bibinfo {author} {\bibfnamefont {B.J.}\ \bibnamefont {Geurts}},\
  }\bibfield  {title} {\enquote {\bibinfo {title} {Heat flux intensification by
  vortical flow localization in rotating convection},}\ }\href@noop {}
  {\bibfield  {journal} {\bibinfo  {journal} {Phys. Rev. E}\ }\textbf {\bibinfo
  {volume} {74}},\ \bibinfo {pages} {056306} (\bibinfo {year}
  {2006})}\BibitemShut {NoStop}%
\bibitem [{\citenamefont {Oresta}\ \emph {et~al.}(2007)\citenamefont {Oresta},
  \citenamefont {Stringano},\ and\ \citenamefont
  {Verzicco}}]{oresta2007transitional}%
  \BibitemOpen
  \bibfield  {author} {\bibinfo {author} {\bibfnamefont {P.}~\bibnamefont
  {Oresta}}, \bibinfo {author} {\bibfnamefont {G.}~\bibnamefont {Stringano}}, \
  and\ \bibinfo {author} {\bibfnamefont {R.}~\bibnamefont {Verzicco}},\
  }\bibfield  {title} {\enquote {\bibinfo {title} {Transitional regimes and
  rotation effects in rayleigh–bénard convection in a slender cylindrical
  cell},}\ }\href@noop {} {\bibfield  {journal} {\bibinfo  {journal} {European
  J. Mechanics - B}\ }\textbf {\bibinfo {volume} {26}},\ \bibinfo {pages} {1 --
  14} (\bibinfo {year} {2007})}\BibitemShut {NoStop}%
\bibitem [{\citenamefont {Stevens}\ \emph {et~al.}(2013)\citenamefont
  {Stevens}, \citenamefont {Clercx},\ and\ \citenamefont
  {Lohse}}]{stevens2013heat}%
  \BibitemOpen
  \bibfield  {author} {\bibinfo {author} {\bibfnamefont {R.J.A.M.}\
  \bibnamefont {Stevens}}, \bibinfo {author} {\bibfnamefont {H.J.H.}\
  \bibnamefont {Clercx}}, \ and\ \bibinfo {author} {\bibfnamefont
  {D.}~\bibnamefont {Lohse}},\ }\bibfield  {title} {\enquote {\bibinfo {title}
  {Heat transport and flow structure in rotating rayleigh–bénard
  convection},}\ }\href@noop {} {\bibfield  {journal} {\bibinfo  {journal}
  {European J. Mechanics - B}\ }\textbf {\bibinfo {volume} {40}},\ \bibinfo
  {pages} {41 -- 49} (\bibinfo {year} {2013})}\BibitemShut {NoStop}%
\bibitem [{\citenamefont {Julien}\ \emph {et~al.}(1996)\citenamefont {Julien},
  \citenamefont {Legg}, \citenamefont {McWilliams},\ and\ \citenamefont
  {Werne}}]{julien1996rapidly}%
  \BibitemOpen
  \bibfield  {author} {\bibinfo {author} {\bibfnamefont {K.}~\bibnamefont
  {Julien}}, \bibinfo {author} {\bibfnamefont {S.}~\bibnamefont {Legg}},
  \bibinfo {author} {\bibfnamefont {J.}~\bibnamefont {McWilliams}}, \ and\
  \bibinfo {author} {\bibfnamefont {J.}~\bibnamefont {Werne}},\ }\bibfield
  {title} {\enquote {\bibinfo {title} {Rapidly rotating turbulent
  rayleigh-b{\'e}nard convection},}\ }\href@noop {} {\bibfield  {journal}
  {\bibinfo  {journal} {J. Fluid Mech.}\ }\textbf {\bibinfo {volume} {322}},\
  \bibinfo {pages} {243} (\bibinfo {year} {1996})}\BibitemShut {NoStop}%
\bibitem [{\citenamefont {Hart}\ and\ \citenamefont
  {Ohlsen}(1999)}]{hart1999thermal}%
  \BibitemOpen
  \bibfield  {author} {\bibinfo {author} {\bibfnamefont {J.E.}\ \bibnamefont
  {Hart}}\ and\ \bibinfo {author} {\bibfnamefont {D.R.}\ \bibnamefont
  {Ohlsen}},\ }\bibfield  {title} {\enquote {\bibinfo {title} {On the thermal
  offset in turbulent rotating convection},}\ }\href@noop {} {\bibfield
  {journal} {\bibinfo  {journal} {Phys. Fluids}\ }\textbf {\bibinfo {volume}
  {11}},\ \bibinfo {pages} {2101--2107} (\bibinfo {year} {1999})}\BibitemShut
  {NoStop}%
\bibitem [{\citenamefont {Liu}\ and\ \citenamefont
  {Ecke}(2011)}]{liu2011local}%
  \BibitemOpen
  \bibfield  {author} {\bibinfo {author} {\bibfnamefont {Y.}~\bibnamefont
  {Liu}}\ and\ \bibinfo {author} {\bibfnamefont {R.E.}\ \bibnamefont {Ecke}},\
  }\bibfield  {title} {\enquote {\bibinfo {title} {Local temperature
  measurements in turbulent rotating rayleigh-b{\'e}nard convection},}\
  }\href@noop {} {\bibfield  {journal} {\bibinfo  {journal} {Phys. Rev. E}\
  }\textbf {\bibinfo {volume} {84}},\ \bibinfo {pages} {016311} (\bibinfo
  {year} {2011})}\BibitemShut {NoStop}%
\bibitem [{\citenamefont {Hakim}\ and\ \citenamefont
  {Canavan}(2005)}]{hakim2005observed}%
  \BibitemOpen
  \bibfield  {author} {\bibinfo {author} {\bibfnamefont {G.J.}\ \bibnamefont
  {Hakim}}\ and\ \bibinfo {author} {\bibfnamefont {A.K.}\ \bibnamefont
  {Canavan}},\ }\bibfield  {title} {\enquote {\bibinfo {title} {Observed
  cyclone–anticyclone tropopause vortex asymmetries},}\ }\href@noop {}
  {\bibfield  {journal} {\bibinfo  {journal} {Journal of the Atmospheric
  Sciences}\ }\textbf {\bibinfo {volume} {62}},\ \bibinfo {pages} {231--240}
  (\bibinfo {year} {2005})}\BibitemShut {NoStop}%
\bibitem [{\citenamefont {Godeferd}\ and\ \citenamefont
  {Moisy}(2015)}]{godeferd2015structure}%
  \BibitemOpen
  \bibfield  {author} {\bibinfo {author} {\bibfnamefont {F.S.}\ \bibnamefont
  {Godeferd}}\ and\ \bibinfo {author} {\bibfnamefont {F.}~\bibnamefont
  {Moisy}},\ }\bibfield  {title} {\enquote {\bibinfo {title} {Structure and
  dynamics of rotating turbulence: a review of recent experimental and
  numerical results},}\ }\href@noop {} {\bibfield  {journal} {\bibinfo
  {journal} {Appl. Mech. Rev.}\ }\textbf {\bibinfo {volume} {67}},\ \bibinfo
  {pages} {030802} (\bibinfo {year} {2015})}\BibitemShut {NoStop}%
\bibitem [{\citenamefont {Bartello}\ \emph {et~al.}(1994)\citenamefont
  {Bartello}, \citenamefont {Metais},\ and\ \citenamefont
  {Lesieur}}]{bartello1994rotating}%
  \BibitemOpen
  \bibfield  {author} {\bibinfo {author} {\bibfnamefont {P.}~\bibnamefont
  {Bartello}}, \bibinfo {author} {\bibfnamefont {O.}~\bibnamefont {Metais}}, \
  and\ \bibinfo {author} {\bibfnamefont {M.}~\bibnamefont {Lesieur}},\
  }\bibfield  {title} {\enquote {\bibinfo {title} {Coherent structures in
  rotating three-dimensional turbulence},}\ }\href@noop {} {\bibfield
  {journal} {\bibinfo  {journal} {J. Fluid Mech.}\ }\textbf {\bibinfo {volume}
  {273}},\ \bibinfo {pages} {1--29} (\bibinfo {year} {1994})}\BibitemShut
  {NoStop}%
\bibitem [{\citenamefont {Deusebio}\ \emph {et~al.}(2014)\citenamefont
  {Deusebio}, \citenamefont {Boffetta}, \citenamefont {Lindborg},\ and\
  \citenamefont {Musacchio}}]{deusebio2014dimensional}%
  \BibitemOpen
  \bibfield  {author} {\bibinfo {author} {\bibfnamefont {E.}~\bibnamefont
  {Deusebio}}, \bibinfo {author} {\bibfnamefont {G.}~\bibnamefont {Boffetta}},
  \bibinfo {author} {\bibfnamefont {E.}~\bibnamefont {Lindborg}}, \ and\
  \bibinfo {author} {\bibfnamefont {S.}~\bibnamefont {Musacchio}},\ }\bibfield
  {title} {\enquote {\bibinfo {title} {Dimensional transition in rotating
  turbulence},}\ }\href@noop {} {\bibfield  {journal} {\bibinfo  {journal}
  {Phys. Rev. E}\ }\textbf {\bibinfo {volume} {90}},\ \bibinfo {pages} {023005}
  (\bibinfo {year} {2014})}\BibitemShut {NoStop}%
\bibitem [{\citenamefont {Gallet}\ \emph {et~al.}(2014)\citenamefont {Gallet},
  \citenamefont {Campagne}, \citenamefont {Cortet},\ and\ \citenamefont
  {Moisy}}]{gallet2014scale}%
  \BibitemOpen
  \bibfield  {author} {\bibinfo {author} {\bibfnamefont {B.}~\bibnamefont
  {Gallet}}, \bibinfo {author} {\bibfnamefont {A.}~\bibnamefont {Campagne}},
  \bibinfo {author} {\bibfnamefont {P.P.}\ \bibnamefont {Cortet}}, \ and\
  \bibinfo {author} {\bibfnamefont {F.}~\bibnamefont {Moisy}},\ }\bibfield
  {title} {\enquote {\bibinfo {title} {Scale-dependent cyclone-anticyclone
  asymmetry in a forced rotating turbulence experiment},}\ }\href@noop {}
  {\bibfield  {journal} {\bibinfo  {journal} {Phys. Fluids}\ }\textbf {\bibinfo
  {volume} {26}},\ \bibinfo {pages} {035108} (\bibinfo {year}
  {2014})}\BibitemShut {NoStop}%
\bibitem [{\citenamefont {Naso}(2015)}]{naso2015cyclone}%
  \BibitemOpen
  \bibfield  {author} {\bibinfo {author} {\bibfnamefont {A.}~\bibnamefont
  {Naso}},\ }\bibfield  {title} {\enquote {\bibinfo {title}
  {Cyclone-anticyclone asymmetry and alignment statistics in homogeneous
  rotating turbulence},}\ }\href@noop {} {\bibfield  {journal} {\bibinfo
  {journal} {Phys. Fluids}\ }\textbf {\bibinfo {volume} {27}},\ \bibinfo
  {pages} {035108} (\bibinfo {year} {2015})}\BibitemShut {NoStop}%
\end{thebibliography}%

\end{document}